\documentclass[letterpaper, 11pt, oneside, notitlepage, twocolumn, ]{article}

\usepackage[margin=0.5in, bottom=0.75in, top=0.5in]{geometry}
\usepackage{graphicx, lipsum, textcomp, wrapfig, amsmath}
\usepackage{hyperref}
\usepackage[utf8]{inputenc}
\usepackage{comment}
\usepackage{subcaption}
\usepackage{qrcode}
\usepackage{xcolor}

\usepackage[switch,columnwise]{lineno}

\usepackage[frozencache,cachedir=.]{minted}

\usepackage{abstract}
\usepackage{titling}

\usepackage{fancyhdr,graphicx,lastpage}
\definecolor{mypurple}{RGB}{140,54,140}
\definecolor{homered}{RGB}{127, 0, 10}
\definecolor{officeorange}{RGB}{204, 75, 0}
\definecolor{mauroblue}{RGB}{53, 48, 217}
\definecolor{citegreen}{RGB}{15, 133, 13}
\definecolor{hyperlinkpurple}{RGB}{42, 0, 163}
\hypersetup{
    colorlinks=true,
    linkcolor=hyperlinkpurple,
    filecolor=mypurple,      
    urlcolor=teal,
    citecolor=citegreen
}

\usepackage[scaled]{helvet}
 
\usepackage[T1]{fontenc}
\usepackage[scaled]{beramono}
\definecolor{darkgreen}{rgb}{0.05, 0.3, 0.1}
\let\oldtexttt\texttt
\renewcommand{\texttt}[1]{\oldtexttt{\textcolor{darkgreen}{#1}}}

\usepackage{tocbasic}
\DeclareTOCStyleEntry[
  beforeskip=.2em plus 1pt,
  pagenumberformat=\textbf
]{tocline}{section}
\DeclareTOCStyleEntry[
  indent=0.5em,
  numwidth=2em
]{tocline}{subsection}

\usepackage[skip=2pt]{caption}
\def \mauro#1 {\textcolor{mauroblue}{#1}}

\usepackage{fontawesome5}
\newcommand{\orcid}[1]{\href{https://orcid.org/#1}{\textcolor[HTML]{A6CE39}{\small \faOrcid}}}

\newcommand{\github}[1]{\href{https://github.com/#1}{\textcolor[HTML]{70726E}{\small \faGithub}}}

\usepackage[
    style=ieee, 
    isbn=false, 
    url=true, 
    natbib=true, 
    backend=bibtex,
    maxcitenames=1,
    mincitenames=1
    ]{biblatex}

\title{
\vspace{-24pt}
\textbf{Efficient generation of grids and traversal graphs in compositional spaces towards exploration and path planning}
}

\author{
\textbf{
Adam M. Krajewski~\textsuperscript{a}} \orcid{0000-0002-2266-0099} \github{amkrajewski}, 
Allison M. Beese~\textsuperscript{a}
\orcid{0000-0002-7022-3387},
Wesley F. Reinhart~\textsuperscript{a, b}
\orcid{0000-0001-7256-2123},
Zi-Kui Liu~\textsuperscript{a}
\orcid{0000-0003-3346-3696}\\
\footnotesize{
a. Department of Materials Science and Engineering, The Pennsylvania State University, USA}\\
\footnotesize{b.  Institute for Computational and Data Sciences, The Pennsylvania State University, USA}\\
\footnotesize{Corresponding Author: A. M. Krajewski, 216-456-1534, ak@psu.edu}
}

\date{\today}
\renewcommand{\maketitlehookd}{%

\begin{abstract}
    \vspace{-12pt}
    Many disciplines of science and engineering deal with problems related to compositions, ranging from chemical compositions in materials science to portfolio compositions in economics. They exist in non-Euclidean simplex spaces, causing many standard tools to be incorrect or inefficient, which is significant in combinatorically or structurally challenging spaces exemplified by Compositionally Complex Materials (CCMs) and Functionally Graded Materials (FGMs). This work explores such problem spaces conceptually, quantifies their computational feasibility, and implements several essential methods specific to simplex spaces through a high-performance open-source library \texttt{nimplex}. Most significantly, we derive and implement an algorithm for constructing a novel n-dimensional simplex graph data structure, which contains all discretized compositions and all possible neighbor-to-neighbor transitions as pointer arrays. Critically, no distance or neighborhood calculations are performed, instead leveraging pure combinatorics and the ordering in procedurally generated simplex grids, keeping the algorithm $\mathcal{O}(N)$, enabling construction of graphs with billions of transitions in seconds. Furthermore, we demonstrate how such graph representations can be combined to express path-planning problem spaces and to incorporate prior knowledge while keeping the problem space homogeneous. This enables efficient deployment of existing high-performance gradient descent, graph traversal, and other optimization algorithms.
    \vspace{-6pt}
    \rule{0.92\textwidth}{0.5pt}
\end{abstract}
}

\begin{document}

\maketitle

\tableofcontents 


\section{Introduction}

\subsection{Compositional Spaces}

The term \emph{composition} refers to \emph{a} way an entity can be split into a set of distinct components, and it plays a critical role in many disciplines of science, engineering, and mathematics. For instance, in combinatorics, the composition will refer to \emph{a} way a positive integer is split into a sequence of other positive integers. In materials science, chemical composition refers to how a material (or, more generally, matter) is split into distinct components, such as chemical elements, based on considerations such as fraction of atoms, occupied volume, or contributed mass. In economics, portfolio composition may refer to how finite capital is split across assets, such as cash, equity instruments, real estate, and commodities, based on their monetary value.

The definition of a composition will typically allow for the definition of a finite space in which such a composition exists. In the typical case of the composition defined in terms of a sequence of $d$ fractions, such space will be a standard simplex - a $(d-1)$-dimensional polytope of unit length edges defined for points \textbf{x} which satisfy $x_i>0$ and $\sum_{i=0}^d x_i = 1$. Or, in simple terms, the space where all fractions are positive, treated equally, and add up to 1. Some special cases of $d$=2,3,4, corresponding to 1-simplex, 2-simplex, and 3-simplex, are also known as line segment, triangle, and tetrahedron, respectively.

Working within compositional (simplex) spaces requires several additional considerations relative to the more common Euclidean spaces for which most tools were designed. Otherwise, numerous problems can be introduced, ranging from sampling points outside the space, through incorrect density estimates, to incorrect gradient calculations caused by modifying every $x_{j\neq i}$ when changing $x_i$ assumed to be independent.

This work introduces a new high-performance library called \texttt{nimplex} or \textit{NIM library for simPLEX spaces}, created exclusively for working with such spaces efficiently and qualitatively correctly. It was written in low-level Nim language, allowing for careful optimizations without sacrificing readability, and then compiled with a native Python interface for general use. \texttt{nimplex} provides an efficient implementation of (a) several existing methods from literature, discussed in Sections~\ref{ssec:mc}~and~\ref{ssec:fullgrid}), (b) modifications of existing methods discussed in Section~\ref{ssec:internalgrid}), and (c) entirely new capability of representing compositional spaces through graphs enabled by novel algorithm, progressively developed over the course of Section~\ref{sec:simplexgraph}. Furthermore, Section~\ref{ssec:complexes} shows how \texttt{nimplex} can be leveraged to tackle compositional design problems in innovative ways.

Neither compositional spaces nor \texttt{nimplex} are exclusive to any discipline; however, to better showcase its capabilities, two complex, highly-dimensional materials-related problems of high impact are highlighted. At the same time, translating them and their solutions to other areas can be done directly effortlessly and is discussed throughout this work.

\subsection{Compositionally Complex Materials} \label{ssec:compositionallycomplex}

An exemplar of how tackling highly-dimensional problems allows researchers to unlock novel solutions is the class of Compositionally Complex Materials (CCMs), which combine several components (e.g., elements, compounds, or monomers) at significant fractions to take advantage of emergent behaviors, often enabled by the high configurational entropy. CCMs include several sub-classes, such as inorganic Multi Principle Element Alloys (MPEAs) or High Entropy Alloys (HEAs) \cite{Yeh2004NanostructuredOutcomes, Cantor2004MicrostructuralAlloys, Senkov2019HighAlloys, Li2019MechanicalAlloys, Debnath2021GenerativeAlloys}, High Entropy Ceramics (HECs) \cite{Oses2020High-entropyCeramics}, and High Entropy Metallic Glasses (HEMGs) \cite{Wang2014High-EntropyGlasses}, or organic High Entropy Polymers (HEPs) \cite{Huang2021High-entropySeparation, Hou2021Entropy-DrivenPolymers, Hirai2022High-entropyReaction}, and complex biomolecular condensates \cite{Jacobs2021Self-AssemblyComponents}. Furthermore, CCMs are conceptually equivalent to a number of systems found in disciplines seemingly unrelated to materials, such as \textit{polypharmacy} in the context of study of drug interactions \cite{Chou2006TheoreticalStudies, Maher2014ClinicalElderly, Guthrie2015The19952010}, or \textit{complex microbial communities}, which can be synthesized to study microbial dynamics \cite{Wu2024Data-drivenCommunities, vandenBerg2022EcologicalCommunities, vanLeeuwen2023SyntheticApplications}, and methods introduced in this work should be directly applicable to them as well.

Out of all CCMs, HEAs are the most extensively studied class, with thousands of literature publications spanning decades since two pioneering 2004 works by Yeh et al. \cite{Yeh2004NanostructuredOutcomes} and by Cantor et al. \cite{Cantor2004MicrostructuralAlloys}, who independently proposed that equimolar (equal fractions) metallic alloys with more than 5 (Yeh) or between 6 and 9 (Cantor) elements, could form single solid solutions (SSS) thanks to the configurational entropy stabilizing them. Other notable HEA definitions include all alloys with idealized configurational entropy $\Delta S_{conf}  \geq R \ln{5} = 1.61R$ \cite{Li2019MechanicalAlloys} ($\approx2.32$ bits of information in the composition \textbf{x}) or $\Delta S_{conf}  \geq 1R$  \cite{Senkov2019HighAlloys} ($\approx1.44$ bits).

Regardless of the exact class or definition, while individual CCMs contain a few components, they always occupy very high dimensional problem spaces relative to other materials because they are not as restricted in terms of which elements are present. This results in homogeneous datasets often occupying over 30-dimensional spaces (or 10-20 for specific problems, like refractory HEA \cite{Senkov2019HighAlloys}), which are orders of magnitude larger compared to traditional materials with one or two primary elements. This introduces opportunities for finding exceptional candidates in little-explored chemical spaces, as demonstrated by some HEAs with excellent hardness \cite{Senkov2010RefractoryAlloys}, ductility \cite{Zhang2019PrecipitationAlloy}, room temperature strength \cite{Long2019AProperties}, and refractory strength \cite{Senkov2016DevelopmentSuperalloy, Kang2021SuperiorProcess}. 

In recent years, high-throughput thermodynamics-driven combinatorial studies on CCMs have been successfully performed to generate high-performance materials\cite{Elder2023ComputationalValidation, Elder2023ComputationalDown-selection}, utilizing  CALPHAD thermodynamic databases for CCMs/HEAs (e.g., \cite{Ostrowska2020ThermodynamicW, Ostrowska2022ThermodynamicExperiments, Gambaro2024CombinedAlloys}). However, they are often limited to (a) coarse sampling (e.g., spaced at $5/10$at.\% or specific ratios) due to the combinatorial complexity in number of alloys, (b) low-dimensional points (e.g., $d=4$) due to the combinatorial complexity in component interactions tracked in CALPHAD calculations and increasing single evaluation cost \cite{Elder2023ComputationalValidation, Elder2023ComputationalDown-selection}, or sometimes even (c) limited further to particular points such as equimolar alloys \cite{Yan2021AcceleratedLearning}.

To somewhat alleviate these computational cost challenges, ML models started to be used as surrogates for thermodynamic calculations and experiments \cite{Debnath2023ComparingAlloys, Tandoc2023MiningAlloys} or in the role of candidate selection from ML latent space \cite{Rao2022MachineDiscovery}. They are extremely fast relative to traditional methods, usually taking microseconds per prediction, and they may seem to work near-instantly when used as a drop-in replacement. However, when one tries to deploy ML models on more complex systems, the combinatorial complexities involved (discussed in Section~\ref{ssec:combinatorialcomplexities}) may quickly make ML deployment very expensive, prompting optimization of the approach. 

While the ML inference is typically optimized to the computational limits in state-of-the-art tools like \texttt{PyTorch} \cite{Paszke2019PyTorch:Library}, the rest of the customized composition space infrastructure, traditionally designed for thousands of evaluations taking seconds, may become a severe bottleneck when moving to billions of evaluations taking microseconds, as explored throughout this paper. In particular, being able to do the following tasks in nanosecond to microsecond region typically becomes critical and needs to be considered:

\begin{enumerate}

    \item Efficient random sampling from the uniform grids and continuous distributions (Monte Carlo in Section~\ref{ssec:mc}) to facilitate approaches including active learning \cite{Rao2022MachineDiscovery} and generative design \cite{Debnath2021GenerativeAlloys}.
    
    \item Efficient generation of the uniform grids in simplex spaces to facilitate complete screenings, quantitatively explored in Sections~\ref{ssec:fullgrid} and \ref{ssec:internalgrid}.
    
    \item Efficient generation of high-dimensional graph representations with complete connectivity to all adjacent CCM compositions, explored in detail throughout Section~\ref{sec:simplexgraph}, to deterministically allocate problem space structure and facilitate neighborhood-based exploration. This is particularly beneficial for the gradient calculations between neighboring grid points, where one typically has to either (a) na\"ively compute all possible compositional changes despite redundancy (e.g., if at point 1 gradient $+1\%B \atop -1\%A$ from point 1 to 2 and gradient $+1\%C \atop -1\%A$ from point 1 to 3, then at point 2 the gradient $+1\%C \atop -1\%B$ to point 3 can be known) at least doubling the number of required evaluations, or (b) keep track of all visited states through an associative array (dictionary). The latter can, in principle, scale well with the number of visited points ($\mathcal{O}(1)$ avg. time for hash map) but is many times more computationally intensive compared to directly accessing known memory location through a pointer as one can do with a graph data structure.
    
\end{enumerate}

\subsection{Path Planning in Functionally Graded Materials} \label{ssec:functionallygraded}

Another class of materials where complex compositional spaces have to be considered, even if intermediate compositions may not be complex themselves, is the class of Functionally Graded Materials (FGMs), sometimes narrowed to Compositionally Graded Materials (CGMs). In them, a set of compositions is traversed to form a compositional path inside a single physical part in order to spatially leverage combinations of properties that may not be possible or feasible with a homogeneous material \cite{Saleh202030Challenges}. In the simplest binary example, this could mean increasing the porosity fraction as a function of depth from the part surface to achieve a higher performance-to-weight ratio \cite{Chen2023FunctionallyReview}.

The computational design of advanced FGMs enable solutions to otherwise impossible challenges, such as the creation of compositional pathways between stainless steel and titanium alloys to allow for additive manufacturing (AM) of aerospace and nuclear components, combining these alloys within a single print \cite{Bobbio2022DesignCompositions}. Such a task is highly non-trivial as the simple linear mixing causes several brittle or otherwise detrimental Fe-Ti and Cr-Ti phases to form, regardless of how gradual the compositional gradient is \cite{Reichardt2016DevelopmentManufacturing}. Formation of such phases in significant quantities is not specific to this alloy pair; thus, all possible ternary systems in Cr-Fe-Ni-Ti-V space had to be considered and manually arranged together by experts to obtain a pathway navigating through feasible regions \cite{Bobbio2022DesignCompositions}.

While in recent years, the fabrication of FGMs has become dominated by Directed Energy Deposition AM for various geometries (e.g., radial deposition \cite{Hofmann2014DevelopingManufacturing}), several other notable manufacturing techniques allow the deployment of such pathways. These include deposition-based methods for high-precision applications, casting-based methods for high-volume fabrication \cite{Saleh202030Challenges}, and recently, brazing consecutive metallic foils \cite{Wu2023ATechnology} to create relatively thin compositionally graded interfaces on mass.

\begin{figure}[ht!]
    \centering
    \includegraphics[width=0.38\textwidth]{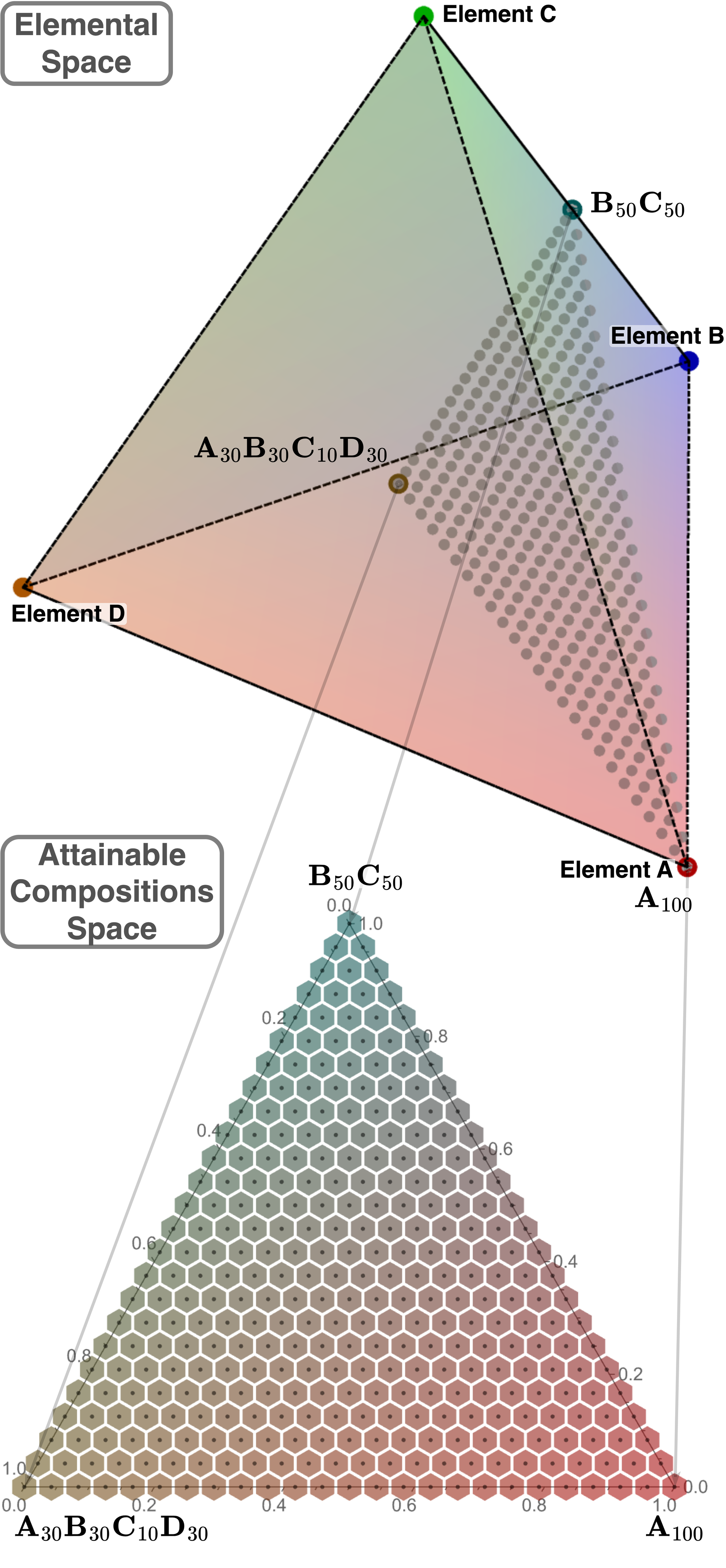}
    \caption{Three available compositions existing in a quaternary (d=4) compositional space forming a ternary (d=3) compositional space which can be attained with them; sampled with a uniform grid with 24 divisions. The hexagonal tiling emerges based on the distance metric in 2-simplex and would become rhombic dodecahedral in 3-simplex.} 
    \label{fig:fgmspaces}
    \vspace{-12pt}
\end{figure}

In a typical FGM manufacturing scenario, a discrete set of compositions (individual available materials) exists in a compositional (simplex) space formed by a union of all components (usually chemical elements or compounds - not affecting later steps), as depicted in the top of Figure~\ref{fig:fgmspaces}, which one could call the \emph{elemental space}. The position in this elemental space is fundamental and is usually the one considered in both mechanistic (e.g., thermodynamic CALPHAD-type models \cite{Olson2023GenomicDynamics}) and predictive (ML/empirical-rule) modeling. However, during the FGM design, it is more convenient to consider another compositional space formed by treating the original available compositions as components, as depicted on the bottom of Figure~\ref{fig:fgmspaces}, which one could call \emph{attainable compositions space} or more generally the \emph{design space}.

Within an FGM manufacturing apparatus, it is common for each of the available compositions to be treated equally, e.g., powder hoppers \cite{Reichardt2021AdvancesMaterials}, sputtering targets \cite{Wu2023ATechnology}, or other flow sources are symmetrically arranged and offer the same degree of control. Thus, as depicted in Figure~\ref{fig:fgmspaces}, the attainable compositional space can be treated as a standard simplex for design purposes and partitioned equally across dimensions, reflecting the nature of the problem even though equidistant points in it may not be equidistant in the original (elemental) space. Notably, a critical consequence of such a setup is the ability to control the relative spacing of the design grid in different dimensions of the elemental space to enable careful control of the element ranges. For instance, one can consider an attainable space constructed from several individual metallic alloys and their equal mixture with 3\% of carbon added, enabling fine control of its content in the design.

The attainable spaces used in the final design tend to be lower-dimensional relative to the corresponding elemental spaces, especially when the available compositions are CCMs/HEAs or the number of flow sources is limited. However, this trend is not fundamentally required, and going against it may be beneficial in many contexts. For instance, one may conceptualize a ternary ($d=3$) elemental compositional space where 4 compositions are available, arranged as vertices of some tetragon; thus, forming a quaternary ($d=4$) attainable compositions space tetrahedron. In such a case, some regions have to overlap in the elemental space, while new regions are guaranteed to be unlocked relative to 3 available compositions if the formed tetragon is strictly convex. This seemingly oversamples; however, it is critical to consider that there is no oversampling in the design space because the available materials can possess properties that are not a function of the composition alone, such as the $CO_2$ footprint or price. 

A clear and industry-significant example of the above happens during FGM design in elemental spaces containing Hf and Zr. The two are very difficult to separate, causing both price and greenhouse emissions to rise sharply as a function of the separation purity requirements. Furthermore, physical form factors available from suppliers tend to be limited or at lower demand for pure-Zr and pure-Hf, furthering the cost. In 
the case of AM using wires as feedstock (WAAM) \cite{Shen2016FabricationProcess}, as explored in detail in \nameref{app1}, using pure Zr in place of the more industry-common alloy with $4.5\%$Hf can be somewhere from a few times to over 100 times more expensive. In a typical, manual FGM design, a researcher selects one of the two grades based on their expertise. However, by considering the two grades as independent components of the higher-dimensional design space, one can avoid forcing a decision before exploring the space, thus limiting human bias and allowing exploration of both options simultaneously, allowing their combination where regions of space insensitive to the Hf content utilize the cheaper grade while the pure Zr is used when necessary or favorable based on some path heuristic.

\begin{figure}[H]
    \centering
    \includegraphics[width=0.475\textwidth]{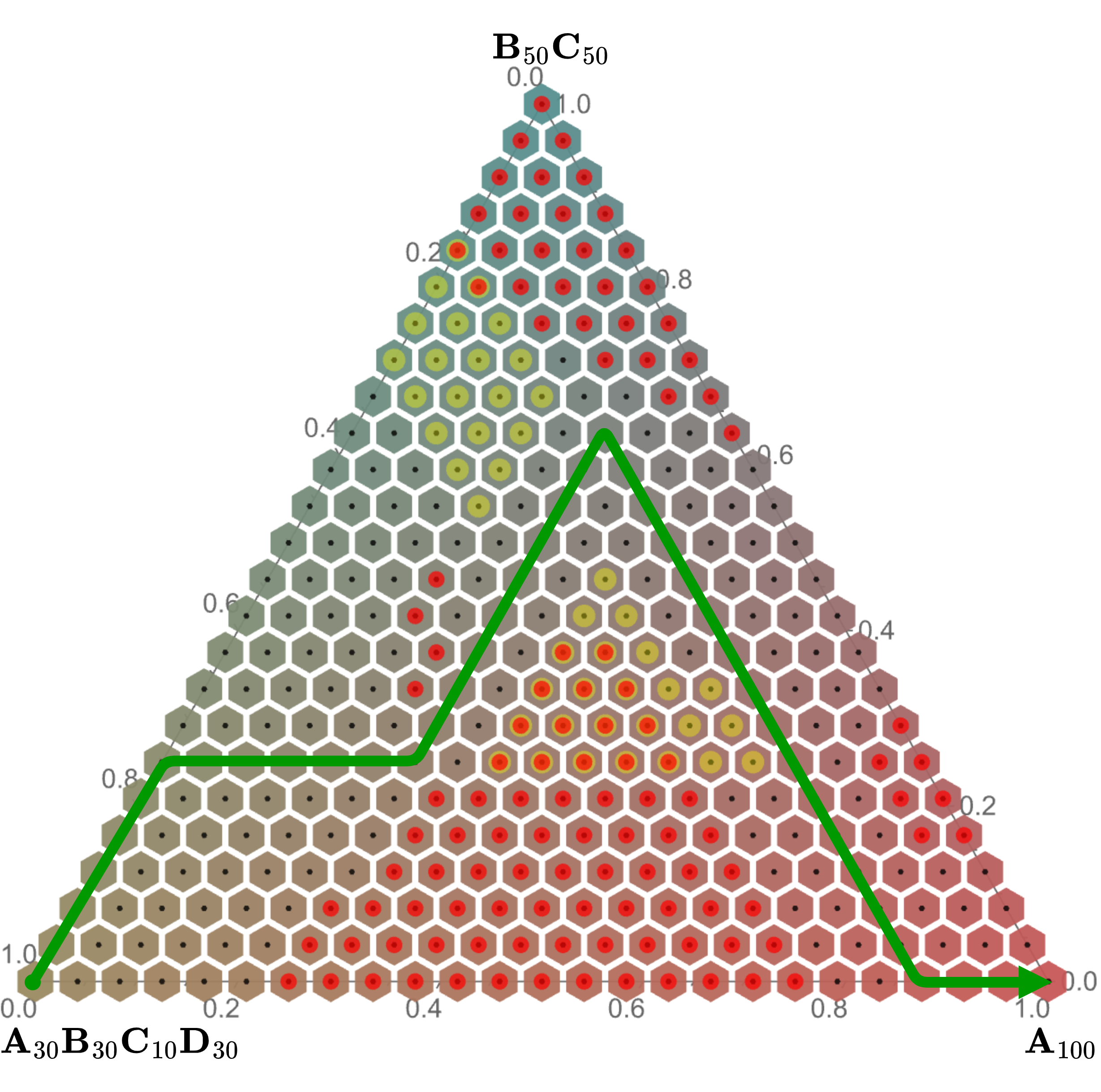}
    \caption{A path example which avoids infeasible (red) and undesirable (yellow) regions, or their combination (orange).} 
    \label{fig:pathplanning1}
\end{figure}

With the design space carefully set up, one can start to evaluate different paths across it. Typically, the core considerations deal with meeting specific feasibility (hard) constraints. In the case of metal alloy FGMs, these can be (1) formation of detrimental phases based on thermodynamic equilibrium \cite{Reichardt2021AdvancesMaterials}, (2) formation of detrimental phases based on non-equilibrium predictions of solidification results based on Scheil–Gulliver method, which better describes the as-made material \cite{Bocklund2020ExperimentalMaterials}, or (3) a combination of the two \cite{Bobbio2022DesignCompositions}. In the future, these will likely be extended through (4) precipitation modeling improving the metastable material design, thanks to the recent release of open-source high-performance software Kawin \cite{Ury2023Kawin:Model}, and (5) automated modeling of manufacturing constraints, such as printability in AM \cite{SheikhAnAlloys}. Furthermore, one can also try to meet desirability (soft) constraints, such as the physical appearance of a composite material, which can be broken if needed. These two types of constraints are depicted in Figure~\ref{fig:pathplanning1}, alongside an example path navigating through them.

\vspace{-6pt}

In Figure~\ref{fig:pathplanning1}, all infeasible points violating the constraints are annotated for visualization. However, doing so may be unnecessary when path-planning, especially iteratively based on neighbor connectivity, as the insides of the infeasible space could not be reached, thus reducing the total number of evaluations.

\begin{figure}[h!]
    \centering
    \includegraphics[width=0.475\textwidth]{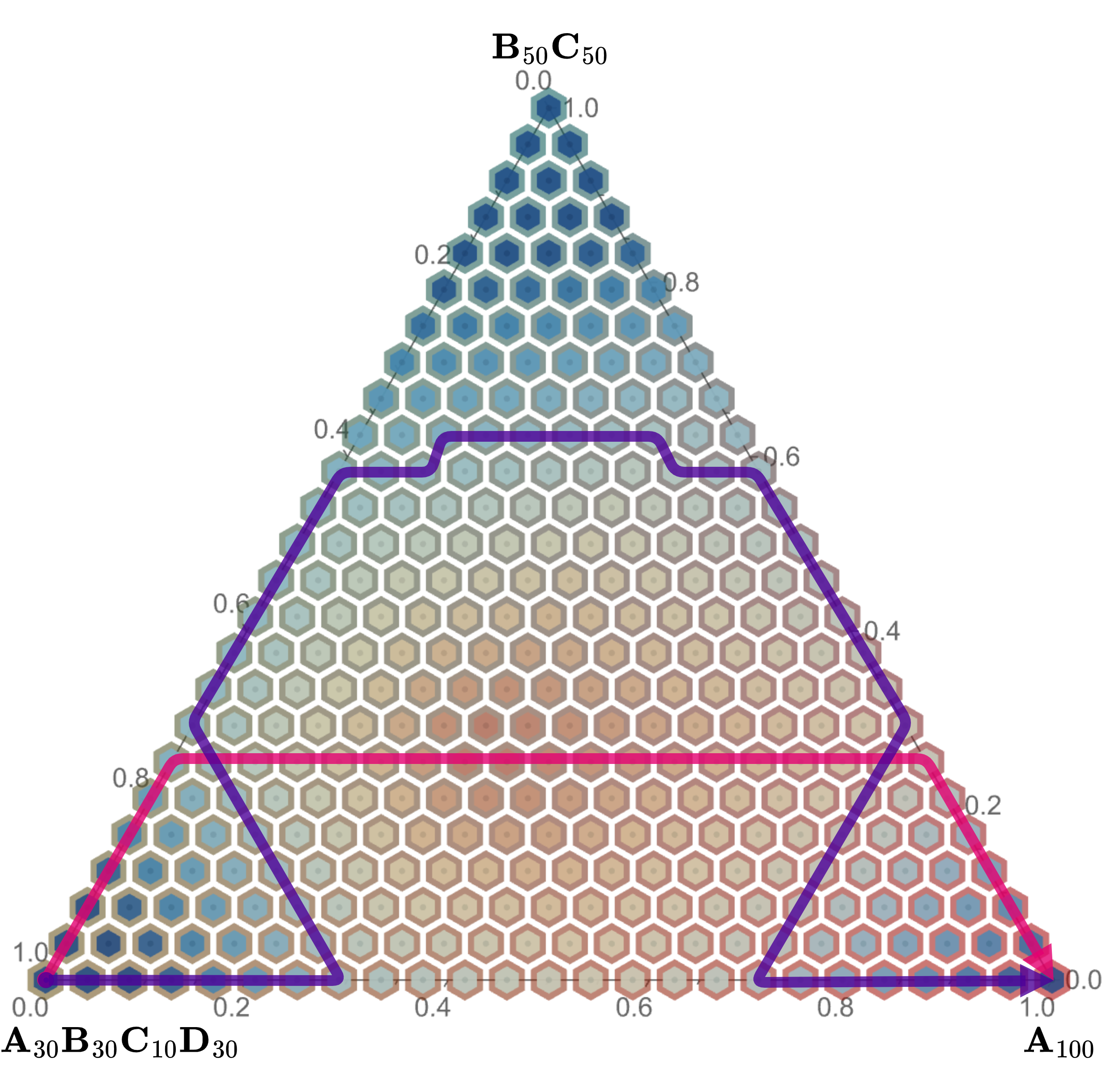}
    \caption{Two path examples in the attainable compositional space overlaid with hypothetical property values, for illustration purposes. One (pink/inner) minimizes/maximizes the average property value given a number of fixed path lengths, and another (purple/outer) minimizes the gradient in the property along the path.} 
    \label{fig:pathplanning2}
\end{figure}

In addition to the feasibility and desirability constraints, further considerations are often made to how the path optimizes values of a set of properties of interest, either individually or through some heuristics combining them. Usually, this optimization constitutes finding the path that minimizes or maximizes either average or extreme values over the set of visited states, exemplified by the pink path in Figure~\ref{fig:pathplanning2}. In the case of metal alloy FGMs, this can mean, for instance, minimizing the average evaporation rate of the molten metal \cite{Mukherjee2016PrintabilityManufacturing}, minimizing the maximum susceptibility to different cracking mechanisms \cite{Yang2023DesignCracking}, or maximizing the ductility \cite{Hu2021ScreeningAlloys}.

\vspace{-6pt}

The last, fundamentally different, property optimization task has to do with the gradient, or more generally, the character, of transitions between intermediate states, which will be critical later in the context of graphs in Section~\ref{sec:simplexgraph}. Most commonly, one optimizes the path to minimize value function gradients, exemplified by the purple path in Figure~\ref{fig:pathplanning2}, in order to, for instance, minimize the thermal expansion coefficient mismatch and by extension stresses induced by temperature changes \cite{Kirk2021ComputationalMonotonicity}.

\subsection{Combinatorial Complexities} \label{ssec:combinatorialcomplexities}

As eluded to in Sections~\ref{ssec:compositionallycomplex} and \ref{ssec:functionallygraded}, when sampling compositions or partitioning corresponding spaces, the resulting combinatorial complexities have to be considered to determine whether a method will be computationally feasible. There are two key equations governing these complexities based on (1) the dimensionality of the space (number of components) $d$ and (2) the number of equal divisions made in each dimension $n_d$, which can be found for every feasible fractional step size (such that it can add to $100\%$). 

The first, very intuitive equation gives the number of samples $N_C$ on a Cartesian grid in $d-1$ dimensions, with $-1$ term due to one of the components being considered dependent.
\begin{equation}
    N_C(d, n_d) = (n_d+1)^{d-1}
    \label{eq:nc}
\end{equation}
The second equation gives the number of ways $n_d$ balls can be arranged in $d$ bins, which is well known to be equivalent to much simpler problems of choosing $d-1$ moves or $n_d$ placements from $d-1+n_d$ possible options (see \cite{Nijenhuis1978CombinatorialCalculators} or \cite{Chasalow1995AlgorithmPoints}). While these may seem unrelated to compositions, the former problem is precisely equivalent to finding a composition of an integer or distributing $n_d$ compositional fractions $\frac{1}{n_d}$ across components or chemical elements, giving the number $N_S$ of unique discrete compositions in the simplex space.
\begin{equation}
    N_S(d, n_d) = \binom{d-1+n_d}{d-1} = \binom{d-1+n_d}{n_d}
    \label{eq:ns1}
\end{equation}
In terms of factorials, both expressions can be simplified to the same
\[N_S(d, n_d) = \frac{(d - 1 + n_d)!}{(d-1)!n_d!}\]
Throughout Sections~\ref{sec:simplexgrid}~and~\ref{sec:simplexgraph}, the interplay between these equations will be utilized to contrast different computational methods, and their direct results will allow computational feasibility evaluation.

\section{Simplex Uniform Random Sampling} \label{sec:randomuniformsampling}

\subsection{Monte Carlo} \label{ssec:mc}

Performing a uniform random sampling, also known as the Monte Carlo method, over a simplex space is a prevalent task; however, it is also one of the most common sources of inefficiency, bias, or errors when implemented incorrectly.

\begin{figure}[h]
    \centering
    \includegraphics[width=0.135\textwidth]{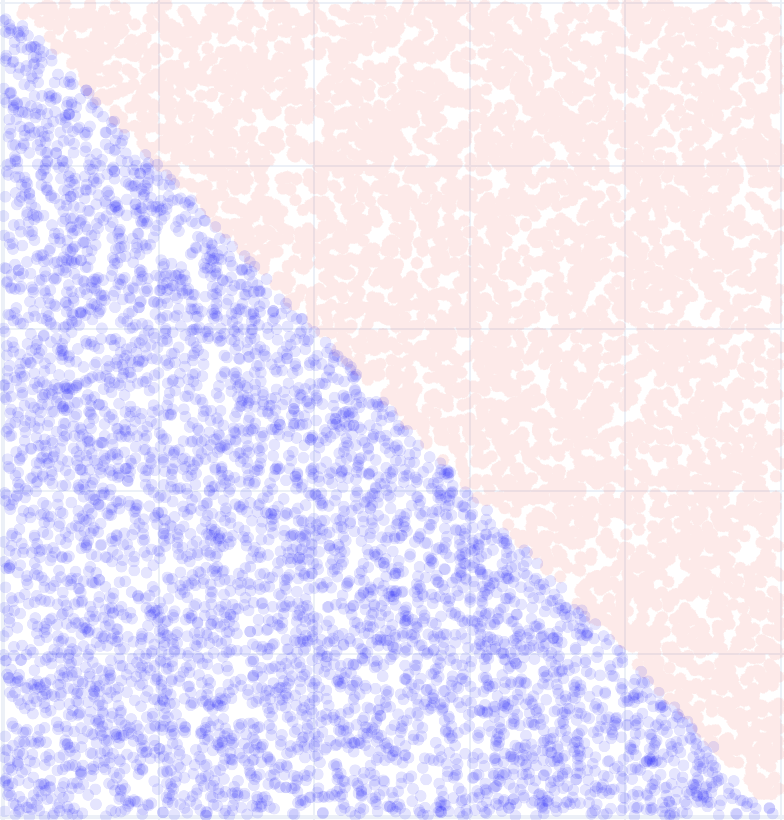}
    \hfill
    \includegraphics[width=0.16\textwidth]{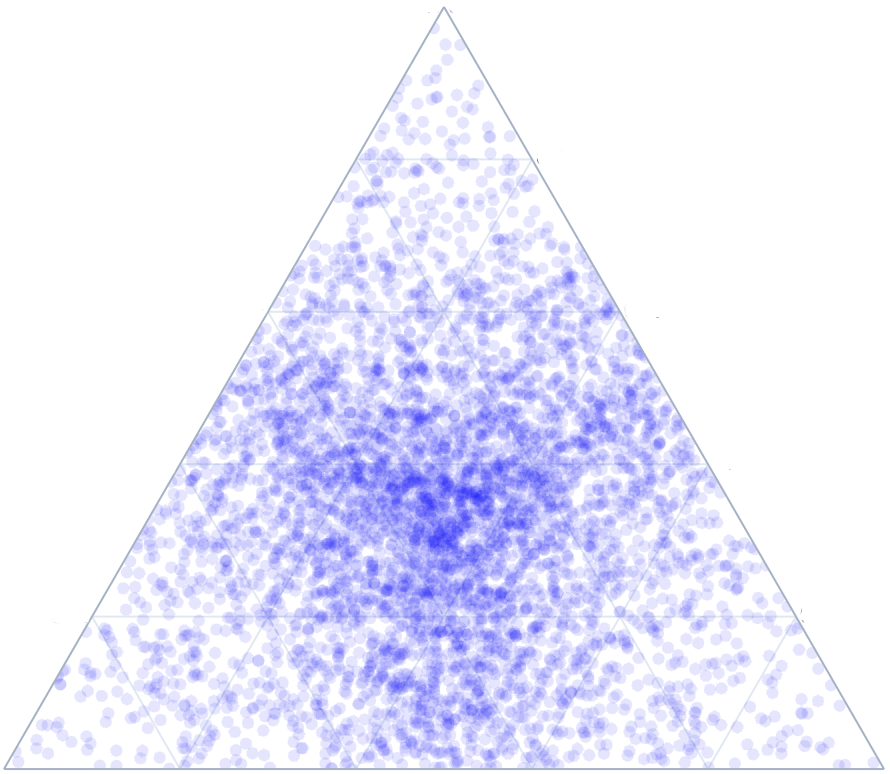}
    \hfill
    \includegraphics[width=0.16\textwidth]{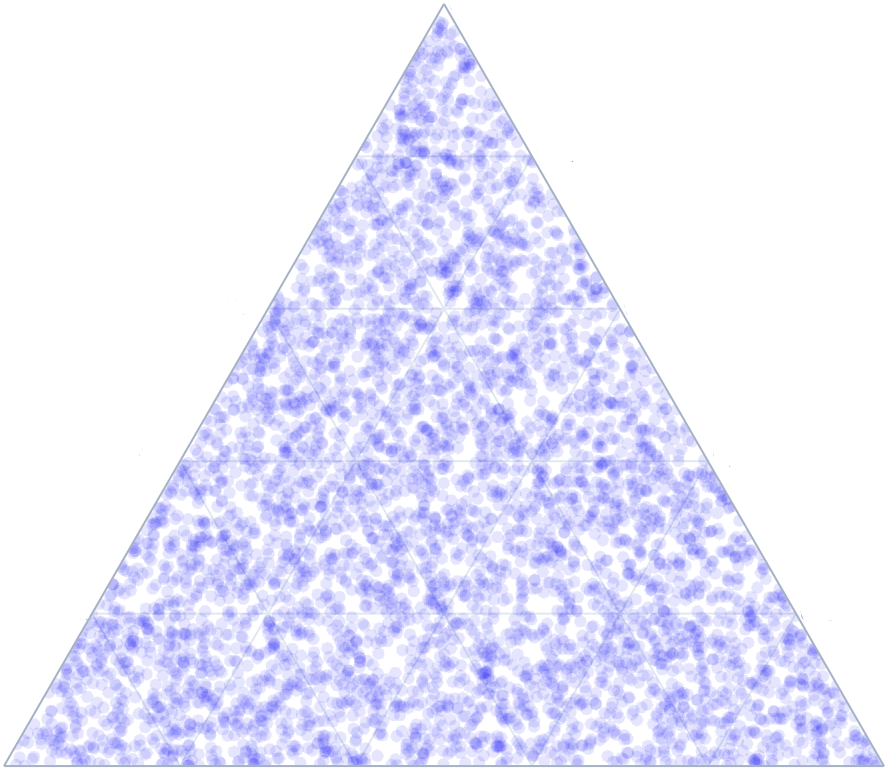}
    \caption{(left) Uniform random sampling in 2-cube (square) filtered to fall onto a 2-simplex (ternary composition), showing 50\% rejection rate, (middle) random sampling in 3-cube projected onto 2-simplex by normalizing coordinates, showing oversampling in the center of each dimension, and (right) ideal uniform random sampling of a simplex.} 
    \label{fig:samplinginternary}
\end{figure}

Software (e.g., \texttt{alchemyst/ternplot} in Matlab \cite{Sandrock2017Alchemyst/ternplothttps://github.com/alchemyst/ternplot}) and methods dealing with low-dimensional or otherwise small compositional spaces, often utilize a na\"ive approach of sampling uniformly distributed points from a Cartesian space/grid in $d-1$ dimensions and then rejecting some infeasible points ($\sum^d_i x_i > 1$), as depicted in the left part of Figure~\ref{fig:samplinginternary}, which for small ($d \leq 4$) can be both easiest and computationally fastest. 

However, this method becomes inefficient for large $d$ because the fraction of rejected points increases with the dimensionality. While this problem is widely noted in the literature \cite{Allen2022AAlloys}, best to the authors' knowledge, it has yet to be discussed quantitatively despite being critical to estimating the sampling's computational complexity. Thus, it is derived herein.

One can consider that a grid of $N_S$ simplex-feasible points is a subset of a grid of $N_C$ points distributed uniformly in the Cartesian space so that random selection from this grid should have a $\frac{N_S}{N_C}$ probability of falling within the simplex. Thus, as shown below, one can find the acceptance rate by considering an infinitely fine grid ($n_d\rightarrow\inf$). \nameref{app2} gives an alternative, intuitive method for finding $f(4)$ using geometry, which agrees with this result.
\begin{equation}
    \begin{aligned}
        f(d) &= \lim_{n_d\rightarrow\inf} \frac{N_S}{N_C} = \lim_{n_d\rightarrow\inf} \frac{\binom{d-1+n_d}{d-1}}{(n_d+1)^{d-1}}\\
        &= \Gamma(d)^{-1} = \frac{1}{(d-1)!} = \frac{d}{d!}
    \end{aligned}
    \label{eq:fd}
\end{equation}
As one can see in Equation~\ref{eq:fd}, the rejection rate exhibits factorial growth, and while it is not a significant obstacle for low-dimensional cases like ternary $f(3)=\frac{1}{2}$ or a quaternary $f(4) = \frac{1}{6}$, it will relatively quickly become a problem when compositionally complex materials are considered. For instance, in the case of nonary chemical space $f(9) = \frac{1}{40320}$ or only $\approx0.0025\%$ of points will fall into the feasible space. Such a rejection rate could have a particularly severe effect on ML-driven methods, such as generative CCM design.

To circumvent the rejection problem, one may randomly sample from N-cube and normalize to 1; however, as shown in the center of Figure~\ref{fig:samplinginternary} and commonly known in the literature \cite{Otis2017AnSystems}, this leads to oversampling in the center of each underlying dimension.

Thus, to achieve a uniform random sampling, \texttt{nimplex} and other carefully designed methods (e.g., \cite{Allen2022AAlloys} and \cite{Otis2017AnSystems}) tend to take Dirichlet distribution, where one samples points $\textbf{y}$ from Gamma distributions with density $\frac{y_i^{\alpha-1} e^{-y_i}}{\Gamma(\alpha)}$ and consider its special "flat" case, where $\alpha=1$ simplifies the density equation to just $\frac{1 e^{-y_i}}{1} = e^{-y_i}$. This is equivalent to sampling $\textbf{z}$ from linear distributions and calculating $y_i=-\log(z_i)$, which then can be normalized to obtain $\textbf{x}$ as $x_i = y_i / \sum \textbf{y}$. The following snippet shows \texttt{nimplex}'s implementation of this, which samples \textbf{z} with the high-performance \texttt{xoroshiro128+} random number generator \cite{Blackman2018ScrambledGenerators} underlying \texttt{randomTensor} function from the \texttt{Arraymancer} tensor library \cite{RatsimbazafyMratsim/Arraymancer:Backends}.

\begin{minted}[linenos=true, mathescape]{nim}
proc simplex_sampling_mc(
    dim: int, samples: int): Tensor[float] =
  let neglograndom = 
    randomTensor[float]([samples, dim], 1.0
    ).map(x => -ln(x))
  let sums = neglograndom.sum(axis=1)
  return neglograndom /. sums
\end{minted}

An alternative approach worth mentioning, sometimes found in this context, is based on (1) generating a $(d+1)$-length list composed of $0$, $d-1$ random numbers, and $1$, (2) sorting it, and (3) obtaining $d$-length list of differences between consecutive elements, which is guaranteed to be uniformly distributed over a simplex as shown in \cite{Rubin1981TheBootstrap}. While this approach may be easier to conceptualize, it is much more computationally expensive due to the sorting step. On the author's laptop, for $d=9$, the method implemented in \texttt{nimplex} (involving calculation of 9 logarithms and normalizing them) takes $3.6$ns while the above (implemented with merge sort) takes $74.5$ns per iteration, i.e., over 20 times longer while not providing any clear benefit. Furthermore, their complexities are $\mathcal{O}(N)$ and $\mathcal{O}(N \ln N)$, respectively, so the computational cost difference will also slowly widen with increasing $d$.

\subsection{Quasi Monte Carlo} \label{ssec:qmc}
While beyond the current implementation scope of \texttt{nimplex}, it is beneficial to consider quasi-Monte Carlo (QMC) sampling methods, where quasi-random sequences of low discrepancy (having highly uniform coverage of all regions) are used to sample the space deterministically. Such an approach is guaranteed to be very beneficial in low-dimensional ($d\leq3$) problems and has been implemented in thermodynamic tools, including \texttt{pycalphad} \cite{Otis2017Pycalphad:Python, Otis2017AnSystems} improving sampling of ternary systems. However, the QMC can become problematic as one moves to higher dimensional problems. 

Firstly, the upper discrepancy bounds for QMC quickly increase with increasing $N$, unlike MC, which depends only on the number of samples; thus, MC \textit{can} outperform it (thanks to better guarantees) unless a quickly (often exponentially) growing number of samples is taken (see discussion on p.271 in \cite{Asmussen2007StochasticAnalysis}). Because of this, even for quaternary ($d=4$) spaces, MC may be preferred for a low number of samples, even though QMC, especially with additional scrambling, \textit{can} outperform it, as shown in \cite{Otis2017AnSystems}. 

Another significant problem in QMC is the unequal sampling of different dimensions, which can be very severe in high dimensions (see p.154 in \cite{Lemieux2009MonteSampling}). In addition to causing under-performance in space-filling, such bias, combined with the standard alphabetical ordering of chemical components, can cause systematically worse exploration of, e.g., titanium compared to aluminum in CCMs, just based on their names.

\section{Simplex Grid} \label{sec:simplexgrid}

\subsection{Full} \label{ssec:fullgrid}

\begin{figure}[h]
    \centering
    \hfill
    \includegraphics[width=0.135\textwidth]{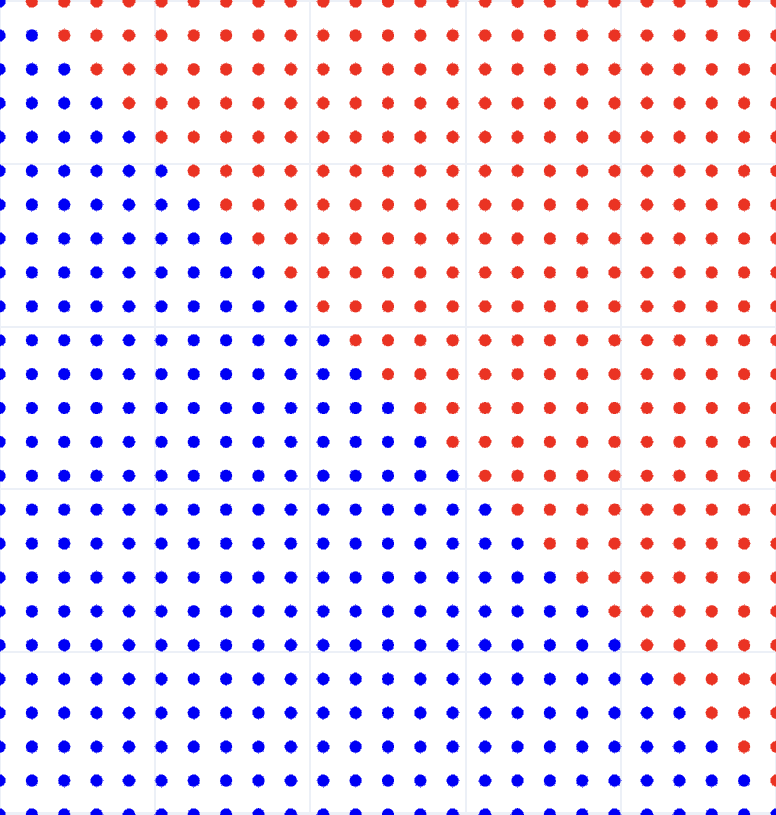}
    \hfill
    \includegraphics[width=0.16\textwidth]{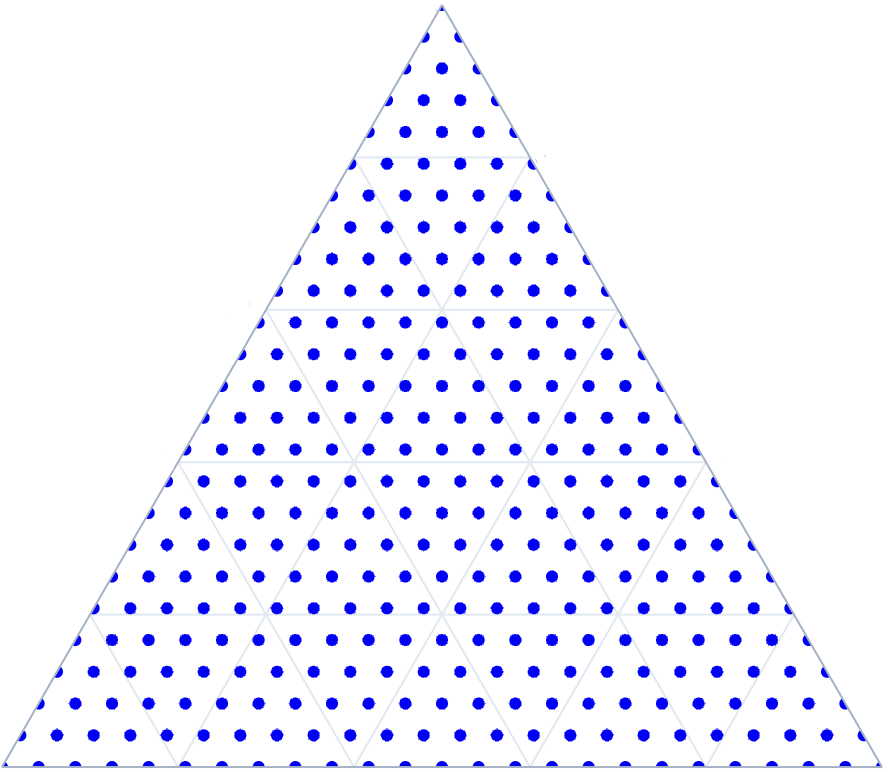}
    \hfill
    \caption{(left) Uniform grid ($n_d=24$) in 2-cube (square) filtered to fall onto a 2-simplex (ternary composition), showing $\frac{12}{25}=48\%$ rejection rate, (right) uniform grid in the corresponding simplex.} 
    \label{fig:simplexgrid}
\end{figure}

Next, one can consider the creation of a grid of uniformly distributed points, which is known to contain $\binom{d-1+n_d}{d-1}$ points, as discussed in Section~\ref{ssec:combinatorialcomplexities}. Similar to the random sampling discussed in Section~\ref{sec:randomuniformsampling}, such a compositional grid cannot be constructed by simply projecting a Cartesian grid in $(N-1)$-cube as patterns will emerge (explored in detail by \citet{Otis2017AnSystems}), but it can be quickly constructed through rejecting infeasible points, as shown in Figure~\ref{fig:simplexgrid}. However, it will suffer from a nearly as bad rejection rate, quantitatively dependent on both $d$ and $n_d$. For instance, if we consider $5\%$ spaced compositions in 9-components, the fraction of points summing to $100\%$ is $f_{M=20}(9) \approx \frac{1}{12,169}$ or $0.0082\%$. 

Fortunately, in their 1978 textbook, \citet{Nijenhuis1978CombinatorialCalculators} explored the problem and gave an efficient algorithm/routine called NEXCOM to procedurally generate these simplex lattice points for arbitrary $d$ and $n_d$, resulting in the grid shown in Figure~\ref{fig:simplexgrid} on the right.

In the following years, several authors made various modifications to the algorithm, and the most recent one by \citet{Chasalow1995AlgorithmPoints} improves performance without sacrificing simplicity. Over the years, it has been implemented in relatively modern languages such as FORTRAN90, C, MATLAB, and Python. Now, it has been implemented in Nim language as well, with the Nim code snippet shown below.

\begin{minted}[linenos=true, mathescape]{nim}
proc simplex_grid(
    dim: int, ndiv: int): Tensor[int] =
  let N: int = binom(ndiv+dim-1, dim-1)
  result = newTensor[int]([N, dim])
  var x = zeros[int](dim)
  x[dim-1] = ndiv
  for j in 0..dim-1:
    result[0, j] = x[j]
  var h = dim
  for i in 1..N-1:
    h -= 1
    let val = x[h]
    x[h] = 0
    x[dim-1] = val - 1
    x[h-1] += 1
    for j in 0..dim-1:
      result[i, j] = x[j]
    if val != 1:
      h = dim
  return result
\end{minted}

As one can deduce from above, the algorithm proceeds through the simplex space starting from $[0, 0, ..., n_d]$ and redistributes one $\frac{1}{n_d}$ fraction $N_S-1$ times across dimensions, forming a zig-zag path to $[n_d, 0, ..., 0]$.

\subsection{Internal} \label{ssec:internalgrid}

To the best of the authors' knowledge, a minor modification that has not been explicitly implemented before, but that is significant to exploration of CCMs  (see Sec \ref{ssec:compositionallycomplex}) is an algorithm to obtain only internal points of the simplex grid, i.e., points with non-zero values in all dimensions, to allow, e.g., generating all 7-component HEAs rather than all alloys in 7-component space. In principle, one can filter the output of the algorithm presented in Section~\ref{ssec:fullgrid}; however, this may quickly become inefficient, especially for $n_d$ low enough as to approach $d$. 

The number of points can be found by, again, considering the surrogate problem of ball compositions mentioned in Section~\ref{ssec:combinatorialcomplexities} and noting that if the last ball cannot be removed from any position, there will be $d$ fewer possible options to perform $d-1$ moves, thus resulting in $N_I$ samples:
\begin{equation}
    N_I(d, n_d) = \binom{n_d-1}{d-1}
\end{equation}
This can be quickly double-checked through summation of internal points of all lower $\delta$ dimensional spaces enclosed in $d$ space:
\[\sum_{\delta=1}^d \Biggl[ \binom{n_d-1}{\delta-1} \times \binom{d}{\delta} \Biggr] =  \frac{(d - 1 + n_d)!}{(d-1)!n_d!} = N_S(d, n_d)\]
We can now look at $N_I(d, n_d)$ to $N_S(d, n_d)$ ratio for the aforementioned case of generating all 7-component alloys. For $5\%$ grid ($n_d=20$) we get $\approx \frac{1}{8.5}$, and for $10\%$ grid ($n_d=10$) we get $\approx \frac{1}{95}$, showing a clear benefit of implementing the new method. This is often done by using the full grid algorithm but with $d$ subtracted from $n_d$, followed by the addition of $1$ to all grid elements, but it can also be accomplished by a dedicated algorithm to avoid any performance overhead. In this work, the latter is done by taking the modified-NEXCOM algorithm \cite{Chasalow1995AlgorithmPoints} from Section~\ref{ssec:fullgrid} and:
\begin{enumerate}
    \item Adjusting procedure length from $N_S$ to $N_I$.
    \item Initializing first states in \textbf{x} to 1.
    \item Adjusting the starting point from [1, 1, ..., $n_{div}$] to [1, 1, ..., $n_{div}-d_{im}+1$].
    \item Jumping to the next dimension one step earlier ($val \neq 2$).
\end{enumerate}

To implement the following \texttt{nimplex} snippet.

\begin{minted}[linenos=true, mathescape]{nim}
proc simplex_internal_grid(
    dim: int, ndiv: int): Tensor[int] =
  let N: int = binom(ndiv-1, dim-1)
  result = newTensor[int]([N, dim])
  var x = ones[int](dim)
  x[dim-1] = ndiv+1-dim
  for j in 0..dim-1:
    result[0, j] = x[j]
  var h = dim
  for i in 1..N-1:
    h -= 1
    let val = x[h]
    x[h] = 1
    x[dim-1] = val - 1
    x[h-1] += 1
    for j in 0..dim-1:
      result[i, j] = x[j]
    if val != 2:
      h = dim
  return result
\end{minted}

\section{Simplex Graph} \label{sec:simplexgraph}

The simplex grid algorithm presented in Section~\ref{ssec:fullgrid} is used commonly; however, it has an important feature that has not been utilized yet and was only briefly noted by its authors \cite{Chasalow1995AlgorithmPoints}. Namely, the fact that generated points are sorted in a lexicographic order (forward or reverse, depending on convention) which opens the door for using pure combinatorics for finding certain interesting relations between points at near-zero costs compared to other popular methods.

\subsection{Binary} \label{ssec:binarygraph}

\begin{figure}[H]
    \centering
    \includegraphics[width=0.45\textwidth]{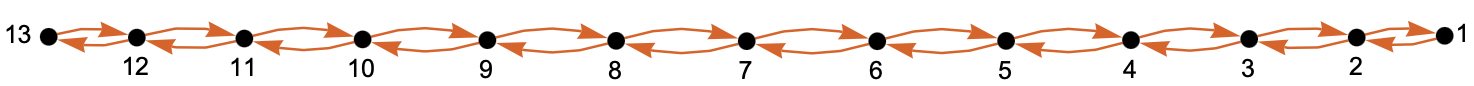}
    \caption{1-simplex graph corresponding to a binary system ($n_d=12$) with 13 nodes/compositions and 24 edges/transitions.} 
    \label{fig:binarysimplexgraph}
\end{figure}

In the simplest possible case, which will be expanded upon later, one can look at a binary ($d=2$ / 1-simplex) compositional grid and write a straightforward function that will find all neighboring points (\emph{transitions} to them) to create a graph representation of the binary system like one presented in Figure~\ref{fig:binarysimplexgraph}, without any notion of distance calculations.

\vspace{-6pt}
\vspace{-6pt}

Such a function, shown below, can be implemented by setting up a \texttt{neighbors} list of lists ($N_S$ of $\leq2$ length) of integer positions and then, at the end of every $i$-th iteration, populating it with forward ($i+1$) and backward ($i-1$) transitions unless start ($[0,1]$) or end ($[1,0]$) points $\textbf{x}$ respectively, corresponding to lack of some component, have been reached.

\vspace{-6pt}
\begin{minted}[linenos=true, mathescape=true]{nim}
proc neighborsLink2C(i:int, x:Tensor, 
    neighbors: var seq[seq[int]]): void =
  if x[0] != 0:
      neighbors[i].add(i+1)
  if x[1] != 0:
      neighbors[i].add(i-1)
\end{minted}

While the above is trivial, it clearly demonstrates that the graph can be constructed within the original $\mathcal{O}(N)$ computational complexity of the simplex grid algorithm, unlike a similarly trivial distance matrix calculation, which would be $\mathcal{O}(N^2)$; thus, unlocking efficient generation of even massive graphs of this kind.

\vspace{-6pt}
\subsection{Ternary} \label{ssec:ternarygraph}

With the core of the approach set up in Section~\ref{ssec:binarygraph}, one can move to the more complex ternary ($d=3$  / 2-simplex) case, which can be conceptualized as a series of $13$ binary systems (already solved individually in Sec. \ref{ssec:binarygraph}) of lengths from $13$ to $1$ and with simple modification of positional coordinates shifted forward by 1 to accommodate for the new dimension. 

The newly allowed neighbor transitions across these binaries can be quickly noticed to be dependent on which of these binaries is considered; however, they can be intuitively found by considering that each transition in the 3rd dimension (increasing $x_0$) limits the size of the binary simplex by 1 from the original size of $\binom{d-1+n_d}{d-1} = \binom{2-1+n_d}{2-1} = n_d+1$. Thus, one can define two convenient jump lengths:
\[
\begin{aligned}
    J_0^{d=3} &= 1\\
    J_1^{d=3}(x_0) &= 1 + n_d - x_0
\end{aligned}
\]
Then, one can quickly visualize that (1) unless $x_2=0$, a transition by jump $J_1$ should be possible, (2) unless $x_1=0$, a transition by jump $J_1$ combined with backward jump $J_0$ in the target binary should be possible, and (3) unless $x_0=0$ (the first traversed binary is considered), transitions by both backward jump $J_1$ and backward jump $J_1+J_0$ (extra step within the earlier binary) should be possible. Thus, one arrives at the following algorithm, which requires additional $n_d$ ("ndiv") input on top of the one from Section~\ref{ssec:binarygraph} but retains its structure.

\begin{minted}[linenos=true, mathescape=true]{nim}
proc neighborsLink3C(...,
    ndiv: int): void =
  let jump0 = 1
  let jump1 = 1+ndiv-x[0]
  if x[0] != 0:
      neighbors[i].add(i-jump1)
      neighbors[i].add(i-jump1-jump0)
  if x[1] != 0:
      neighbors[i].add(i-jump0)
      neighbors[i].add(i+jump1-jump0)
  if x[2] != 0:
      neighbors[i].add(i+jump0)
      neighbors[i].add(i+jump1)
\end{minted}

\begin{figure}[h]
    \centering
    \includegraphics[width=0.45\textwidth]{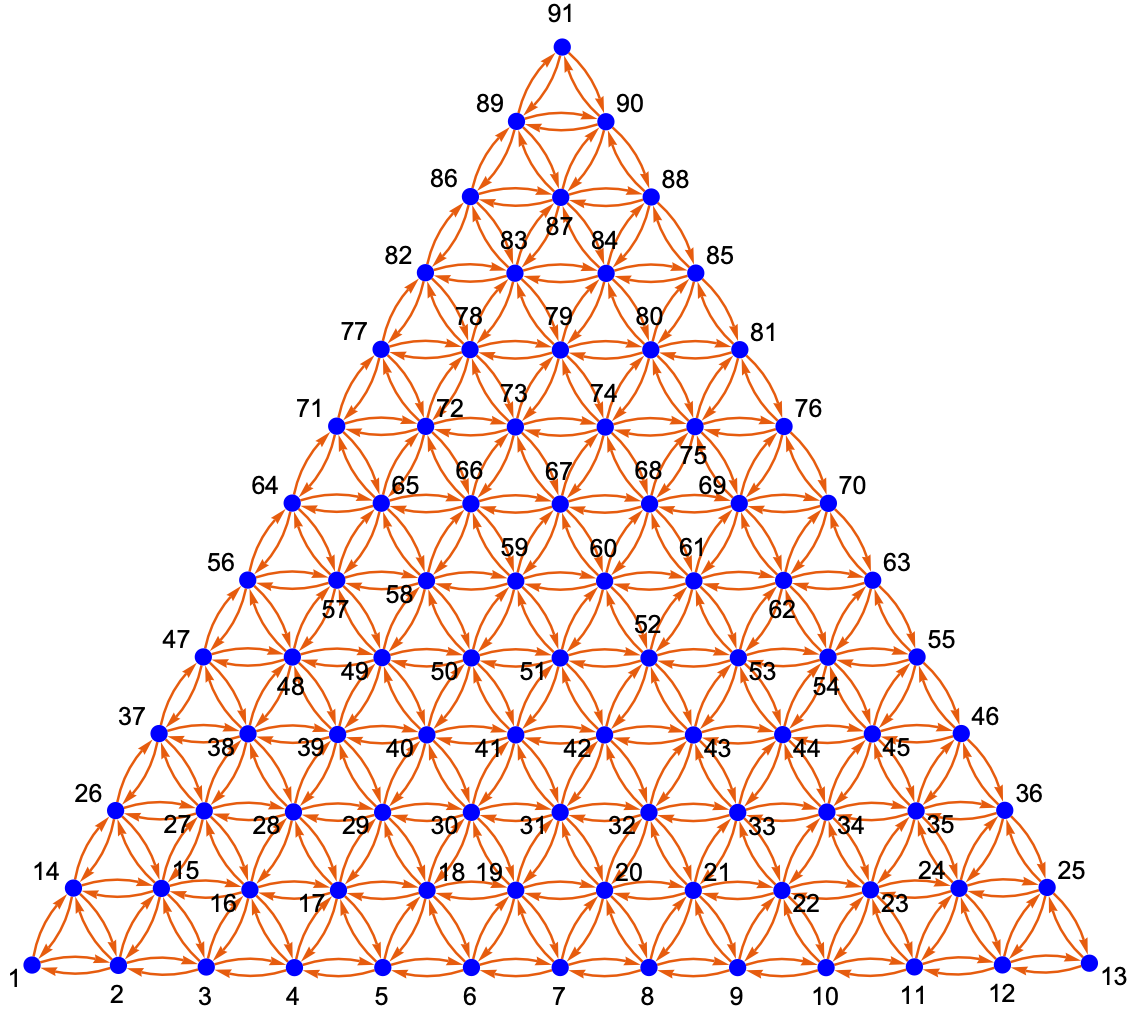}
    \caption{2-simplex graph corresponding to a ternary system ($n_d=12$) with 91 nodes/compositions and 468 edges/transitions.} 
    \label{fig:ternarysimplexgraph}
\end{figure}

Utilizing the above, the result presented in Figure~\ref{fig:ternarysimplexgraph} can be quickly obtained for any number of divisions. The numbering of points can help to visualize how the transitions were obtained.

\vspace{-6pt}

\subsection{N-Dimensional} \label{ssec:ndimgraph}

Moving beyond ternary systems, one has to increase the number of tracked transitions to higher dimensions, which can be counted for every jump length $J_j$ with $\sum_0^{(d-j-2)}x_i$, and then utilized to obtain a general equation for all $d-1$ elements of jump length array $\textbf{J}$ as a function of current point $\textbf{x}$.
\begin{equation}
    J_{j}(\textbf{x})= \binom{j+n_d- \sum_{i=0}^{(d-j-2)} x_i}{j}
    \label{eq:jj}
\end{equation}
As expected, for the special cases of $d=3$, the above agrees with $J_0$ and $J_1$ found for the ternary case in Section~\ref{ssec:ternarygraph}. One can also note that $J_0$ always equals to $1$ as $\binom{a}{0} = 1$ for any $a$.

With $\textbf{J}$ defined, one can take a quaternary system (d=4 / 3-simplex) and perform a similar visualization thought exercise in the head as in Section~\ref{ssec:ternarygraph}, but in 3D, considering the new transitions to 3 neighbors above and 3 neighbors below, in order to set up \texttt{\textbf{neighborsLink4C}} procedure which is presented in \nameref{app3}. 

Such an approach of visualizing and counting the possible jumps in the head becomes (a) challenging for quinary systems (d=5 / 4-simplex) case where one has to visualize 4 forward and 4 backward jumps to and from points inscribed in every tetrahedron formed by the 3-simplex tetrahedral grids, and (b) near impossible for higher orders, both because of the visualization dimensionality and the growing number of neighbors to track, given by $\sum_\delta^d 2(\delta-1)= d(d+1)$ or for $d=$ 6, 7, 8, and 9 corresponding to 30, 42, 56, and 72 neighbors respectively; thus prompting for an alternative.

Fortunately, while performing the above thought exercises for increasing $d$, with transition lengths \textbf{T} expressed as compositions of jump lengths described by \textbf{J}, a careful observer can quickly note that for any dimensionality of the simplex grid, the main challenge in finding the higher-dimensional \textbf{T} lies in distributing the $d-1$ new forward ($x_0$ increment) transitions across all previous $x_i=0$ constraints, while the $d-1$ new backward ($x_0$ decrease) transitions are always possible for $x_0>0$ and follow a relatively simple trend of transition lengths $J_d$, $\sum_{j=d-1}^{d}J_j$, ..., $\sum_{j=0}^{d}J_j$. This allows a relatively simple construction of all backward transitions by stacking them together across all $d-2$ considered dimensions. 

Finally, a simple notion that every backward transition $b \rightarrow a$ of grid point $b$ is associated with a forward transition $a \rightarrow b$ of point $a$ allows for the complete construction of the simplex graph representation of the compositional space.

This is implemented very concisely in the \texttt{nimplex} snippet below, where for every considered dimension $\delta$ from $d$ (highest at $0$th index of $\textbf{x}$) down to $2$ ($(d-2)$th index), the $\delta$ of backward and $\delta$ of forward transitions of lengths $t_k$ are found by iteratively summing jump lengths $J_{\delta}$, $\sum_{j=\delta-1}^{\delta}J_j$, ..., $\sum_{j=0}^{\delta}J_j$, and then used to assign neighborhood.

\begin{minted}[linenos=true, mathescape=true]{nim}
proc neighborsLink(...): void =
  var jumps = newSeq[int](dim-1)
  jumps[0] = 1     #binom(a,0)=1
  for j in 1..<(dim-1):
    jumps[j] = binom(
      j+ndiv-sum(x[0..(dim-2-j)]), j)
  var trans: int
  for order in 0..(dim-2): 
    trans = 0
    if x[order] != 0:
      for dir in 0..(dim-2-order): 
        temp += jumps[dim-2-order-dir]
        neighbors[i].add(i - trans)
        neighbors[i - trans].add(i)           
\end{minted}

\begin{figure}[h]
    \centering
    \includegraphics[width=0.475\textwidth]{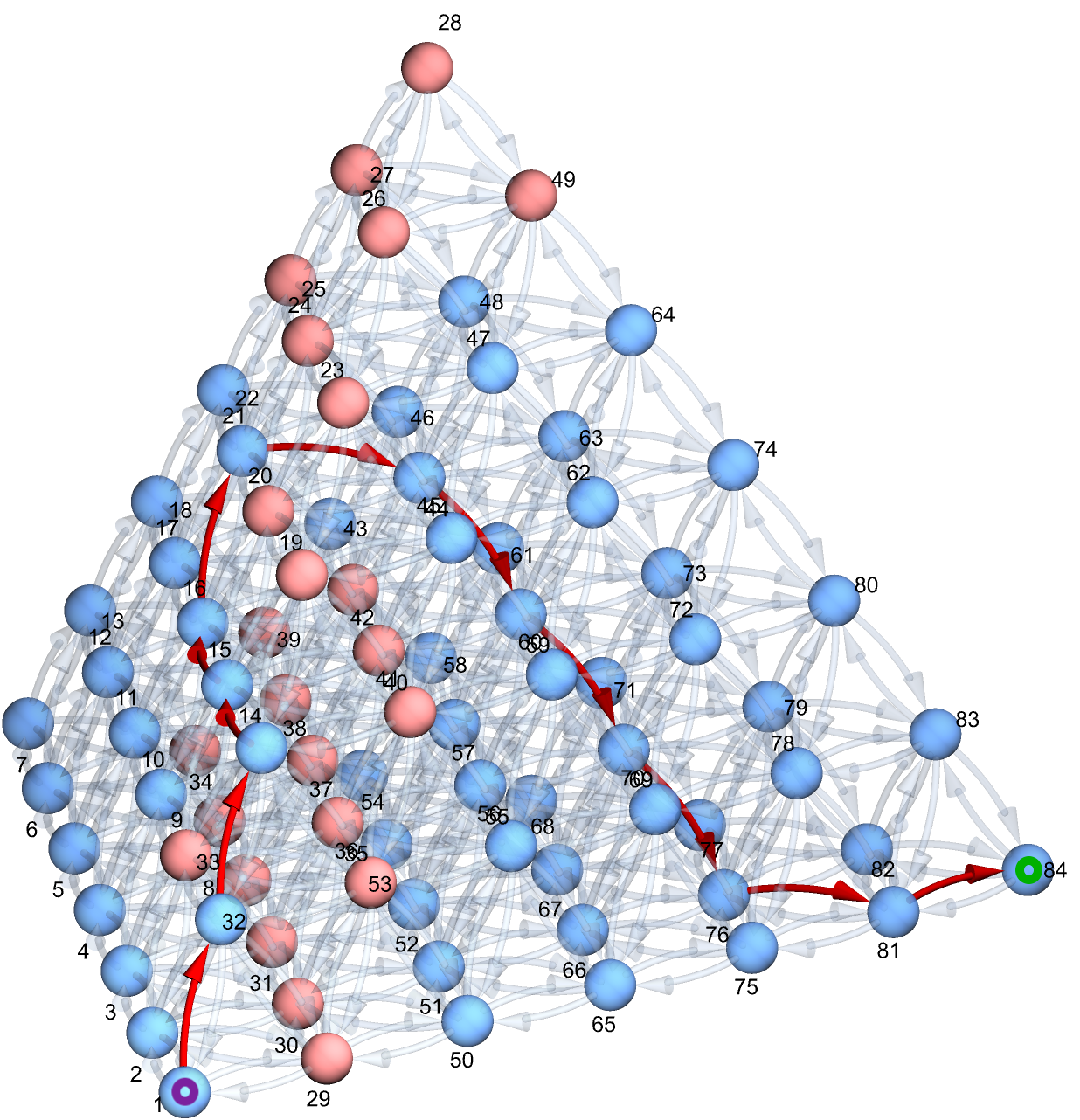}
    \caption{A quaternary (d=4 / 3-simplex) simplex graph ($n_d=6$) with 84 nodes (compositions) and 672 edges (possible moves). A set of nodes has been manually selected (highlighted in pink) to depict a toy example of infeasible points (similarly to Figure~\ref{fig:pathplanning1}), which forces a non-trivial path (highlighted in red) to traverse from the bottom-left corner at 1 to the bottom-right corner at 84.} 
    \label{fig:quaternarysimplexgraph}
\end{figure}

The result of running the above algorithm with $d=4$ and relatively low $n_d$ is shown in Figure~\ref{fig:quaternarysimplexgraph} to help visualize neighbor-neighbor transitions despite the overlap when printed in 2D.

It is critical to note that the above algorithm is still within the $\mathcal{O}(N)$ computational complexity for $N$ grid points, just like the forward/backward jumps discussed in Section~\ref{ssec:binarygraph}. Thus, for instance, the task of constructing 1\% resolution graph for a 6-component chemical space containing $N_S(d=6, n_d=100)$ or nearly \textit{100 million unique vertices} requiring \textit{2.76 billion edges} (possible chemistry changes) takes as little as \textit{23s} tested on author's laptop computer. This stands in stark contrast with $\mathcal{O}(N^2)$ distance-based graph construction, which, even when well implemented to take around $3ns$ per comparison, would take approximately 1 year on the same machine.

Furthermore, the method scales excellently with the increasing problem dimensionality. For a 12-component chemical space with $n_d=12$ divisions per dimension, even though up to $132$ neighbors have to be considered for all $N_S=1.35$ million vertices, the 93 million edges are constructed in 950 milliseconds.

\vspace{-12pt}
\subsection{Simplex Graph Complexes} \label{ssec:complexes}

Once able to rapidly set up simplex graphs in arbitrary dimensions, one can also efficiently combine them to construct more complex graphs representing non-trivial problem statements where many different paths are possible to explore, and prior knowledge can be incorporated as assumptions in the problem solution space if needed. At the same time, it allows the dimensionality of the intermediate compositional spaces to be kept within manufacturing feasibility, i.e., the number of material flow sources.

\begin{figure}[H]
    \centering
    \includegraphics[width=0.35\textwidth]{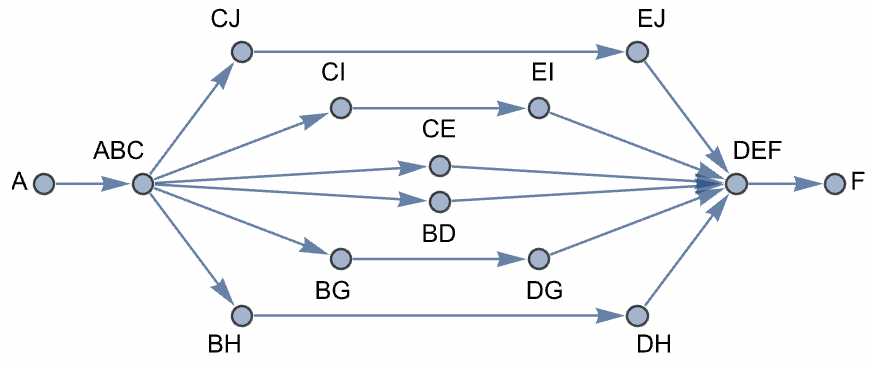}
    \includegraphics[width=0.45\textwidth]{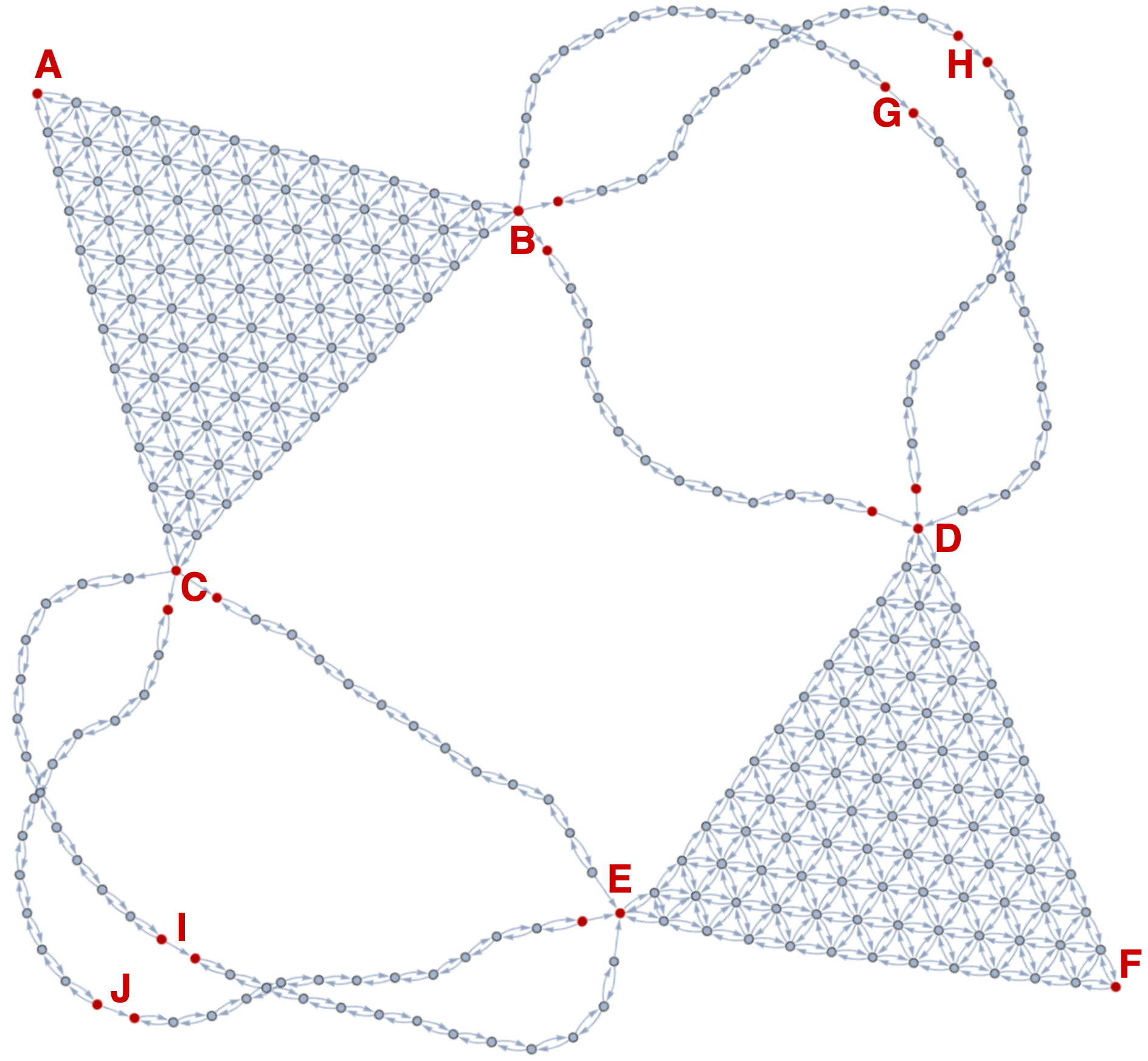}
    \caption{Graph Complex Example \#1 depicting a problem space where 2 ternary systems can be connected through 6 different binary paths.} 
    \label{fig:graphcomplex1}
    \vspace{-12pt}
\end{figure}

Suppose one tries to connect elemental compositions A and F, but assumes prior knowledge that they cannot be combined directly in any quantity, and also knows that (1) A is compatible with B and C, (2) F is compatible with D and E, but (3) B and E are incompatible in any quantity, (4) C and D are incompatible in any quantity. Furthermore, (5) G and H are individually compatible with B and D, and (6) I and J are individually compatible with C and E. These rules can be used to set up a problem graph like in the top of Figure~\ref{fig:graphcomplex1}, encoding everything that is known about the system \textit{a priori} and limiting the solution space from full $\binom{10-1+12}{12} \approx 300,000$ to $2\times\binom{3-1+12}{12} + 10\times\binom{2-1+12}{12} = 312$, or three orders of magnitude.

The space constructed in Figure~\ref{fig:graphcomplex1} is kept very minimal in terms of going beyond known assumptions and dimensionality to illustrate the concept in a plane. However, real examples of this technique can be highly non-trivial and essential in bringing the number of considered points into a computationally feasible regime when tens of available compositions can be considered. 

Furthermore, unlike in Figure~\ref{fig:graphcomplex1} where spaces are simply connected through single-components, the interfaces between the individual compositional spaces can be along any subspace (e.g., the ternary face of quaternary tetrahedron), allowing one to quickly set up search problems where one or more components are unknown, but their relations to others are fixed.

\vspace{-6pt}
\begin{figure}[h!]
    \centering
    \includegraphics[width=0.28\textwidth]{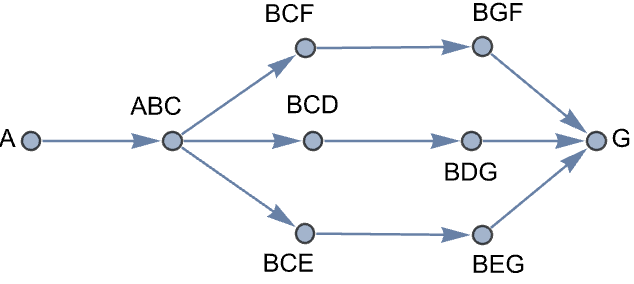}
    \includegraphics[width=0.46\textwidth]{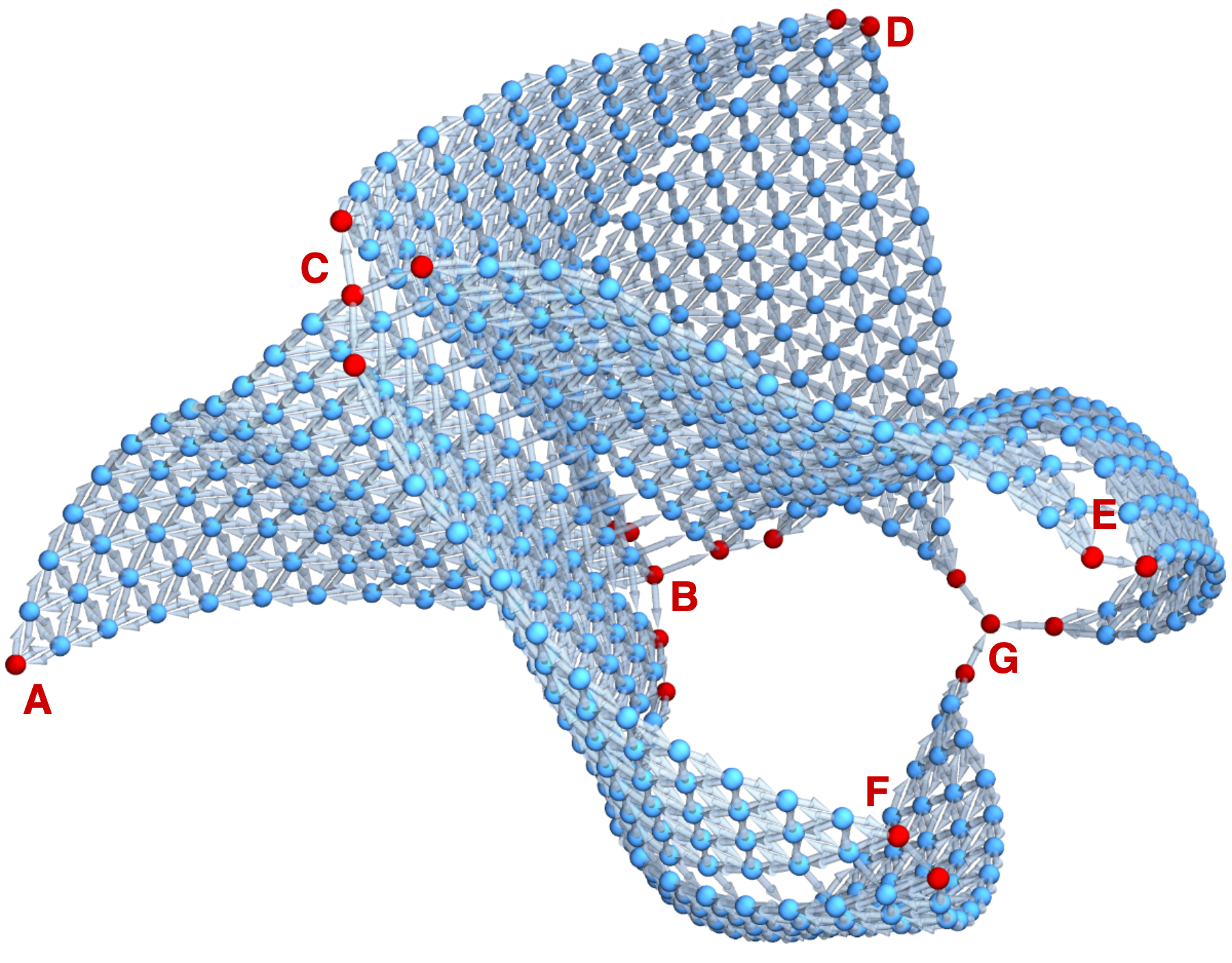}
    \caption{Graph Complex Example \#2 depicting a problem where 3 choices (D/E/F) can be made to traverse from ABC to G through dual ternary systems containing B. Vertices were spread in 3D to depict three possible ABC to G paths, which would exactly overlap in a plane.} 
    \label{fig:graphcomplex2}
    \vspace{-12pt}
\end{figure}

One can quickly demonstrate the benefits of such ability by looking at the SS316 to Ti-6Al-4V problem studied by \citet{Bobbio2022DesignCompositions}. After idealizing and anonymizing the components, it becomes a problem where one tries to combine compositions A with G, which cannot be combined directly in almost any quantity, and also knows that (1) system ABC is highly feasible across it, but (2) C cannot be combined directly with G in any quantity, and (3) a complete path from pure B to G is not possible. In this case, a simple problem setup is to look at several BC\texttt{?} and BG\texttt{?} pairs, forming parallel pathways from ABC to G. This is depicted in Figure~\ref{fig:graphcomplex2} for 3 candidates D, E, F, forming 6 ternary spaces to consider, but nothing limits the method to be extended to an arbitrary number of candidates while still retaining its linear complexity.

\begin{figure}[h]
    \centering
    \includegraphics[width=0.3\textwidth]{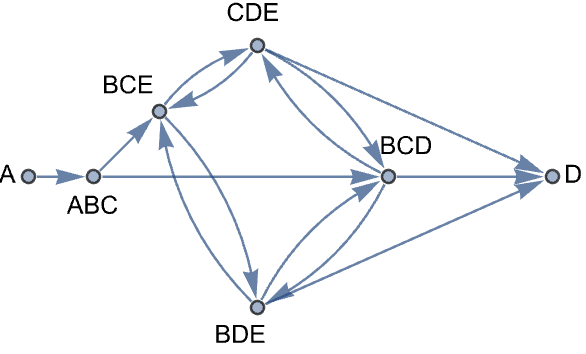}
    \includegraphics[width=0.49\textwidth]{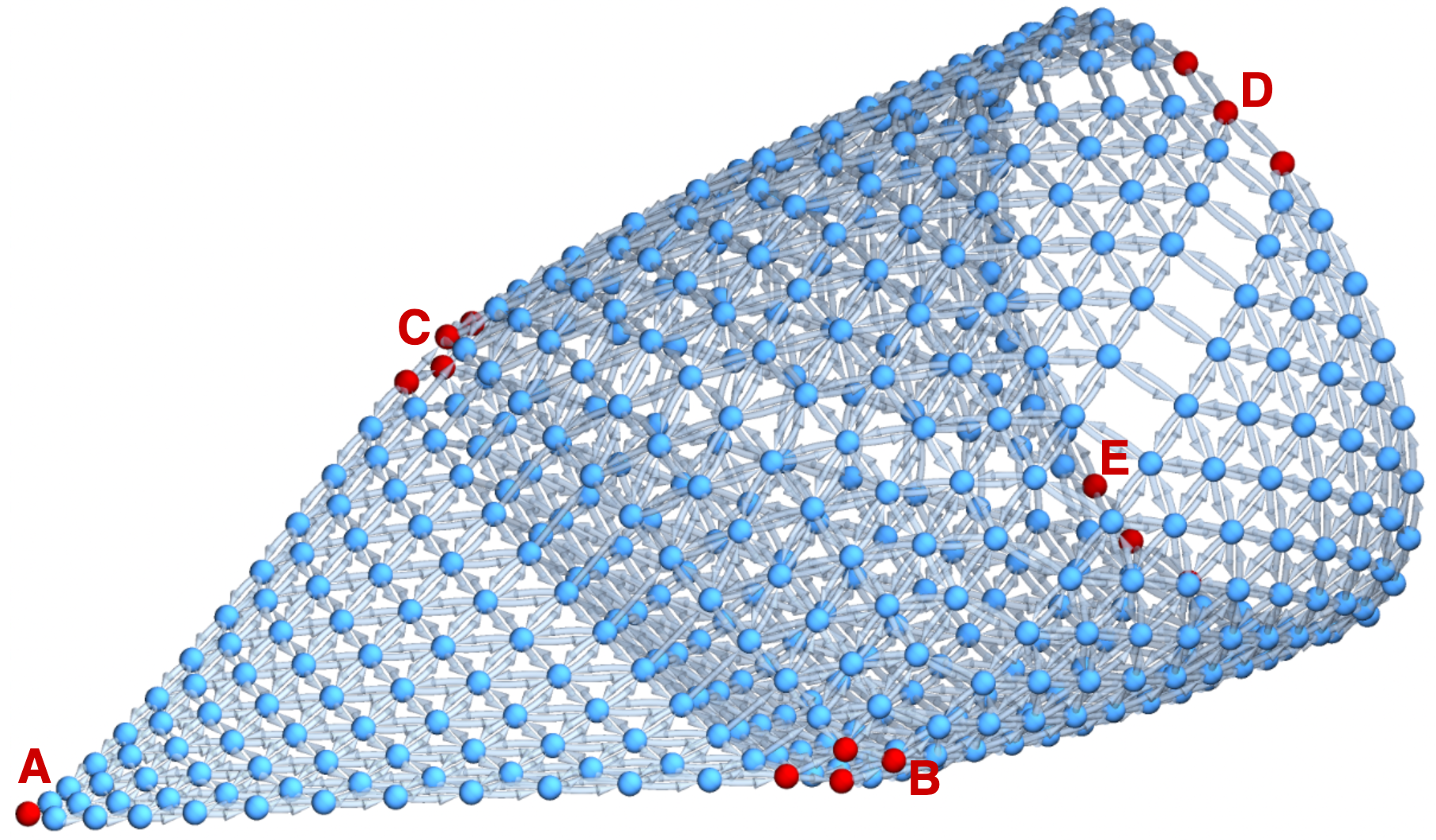}
    \caption{Graph Complex Example \#3 depicting the possibility of competing paths, including cycles.} 
    \label{fig:graphcomplex3}
\end{figure}

In the above examples in Figures~\ref{fig:graphcomplex1}~and~\ref{fig:graphcomplex2}, all connections between compositional spaces were directional; however, that is not necessary, and in some problems it may be beneficial to allow bidirectional movement. Suppose one tries to combine compositions A with D, which cannot be combined directly in any quantity, and also knows that (1) system ABC is highly feasible across it, but (2) system BCD is not traversable on its own. Thus, E can be introduced to set up intermediate spaces BDE and CDE, allowing obstacles in BCD to be avoided. Furthermore, BCE can also be set up as an alternative, possibly shortening the total path. Figure~\ref{fig:graphcomplex3} depicts such a problem setup.

Notably, while the above example in Figure~\ref{fig:graphcomplex3} depicts a single 5th component E to help visualize cycling between spaces, these concepts can be extended to many possible intermediate components. At the same time, the maximum dimensionality of individual compositional spaces is kept constant ($d=3$). Thus, it provides a powerful method to keep the problem solvable, even experimentally, while considering many possible pathways formally defined prior to path planning to fit within the feasibility of evaluation and manufacturing.

\subsection{Exploration} \label{sec:discussion}

Critically, creating such a homogeneous problem structure through graph representation allows one to deploy the same exploration strategies across many dimensionalities and even combinations of individual spaces shown in Section~\ref{ssec:complexes}. Furthermore, in the described graphs, points are on an equidistant grid; thus, it is easy to set up a heuristic function that can be both consistent and admissible. 

This, in turn, enables one to harvest many general-purpose graph traversal algorithms, which are actively researched and available through high-performance libraries. For instance, to navigate against constraints, the $A^*$ algorithm \cite{Hart1968APaths} can be used with such a heuristic and is mathematically guaranteed to find \textit{the} shortest feasible compositional path while exploring the least number of nodes \cite{Dechter1985GeneralizedA}, what can be critical if the shortest path is necessary while each evaluation is relatively expensive. Then, if one tries to find \emph{a} feasible solution first and then improve on it, modifications of the $A^*$ algorithms such as the $RWA^*$ \cite{Bhatia2021OnA} can be used to first make it more greedy and then gradually move towards $A^*$ to obtain the optimal solution if sufficient resources are available. Alternatively, for highly complex problems where exploration needs to proceed quickly towards the goal but the optimality guarantees are not needed, one can use a search algorithm based on the Monte Carlo tree search (MCTS), which has been famously used in conjunction with an ML model to master the game of Go \cite{Silver2016MasteringSearch}.

\section{Discussion} \label{sec:summary}

This work starts by providing an abstract description of compositional spaces applicable to a wide range of disciplines while formalizing several vital concepts. Then, Section~\ref{ssec:compositionallycomplex} discusses complex compositional spaces, using Compositionally Complex Materials (CCMs) as a real-world application and considers the challenges of exploring such spaces using different methods. Section~\ref{ssec:functionallygraded} uses another real-world application of Functionally Graded Materials (FGMs) to expand on that by discussing compositional spaces formed from compositions in other spaces and when these spaces are preferred for design. It also discusses key concepts related to path planning in relation to types of constraint and property optimizations. Last in the Introduction, Section \ref{ssec:combinatorialcomplexities} discusses some equations critical to investigating the combinatorial complexities in these problems.

Next, discussions and implementations are given for several methods for efficiently solving compositional problems through random sampling in Section~\ref{sec:randomuniformsampling}, grid-based methods in Section~\ref{sec:simplexgrid}, graph-based methods, including graphs combining multiple compositional spaces, in Section~\ref{sec:simplexgraph}. The three most critical contributions introduced in this process are:

\begin{enumerate}
    \item Novel algorithm for rapid procedural generation of \emph{N-dimensional graph representations} of compositional spaces where uniformly distributed simplex grid points in $d$ dimensions are completely connected to up to $d(d-1)$ neighbors representing all possible component-pair changes. For instance, in economics, this could represent all possible compositions of a financial portfolio of 12 assets and, for each one of them, all 132 transfer choices that can be made to modify it. Critically, this method scales linearly with the number of points and generates graphs with billions of connections between millions of points in just seconds. Furthermore, this algorithm allows deterministic memory allocation during the graph construction, where arrays of pointers to neighboring compositions represent allowed transitions, resulting in a very high-performance data structure.
    
    \item The new free, open-source software (FOSS) package \texttt{nimplex} (\href{https://nimplex.phaseslab.org}{nimplex.phaseslab.org}), which gives high-performance implementations of both essential existing methods and all new methods introduced within this work including the simplex graphs.
    
    \item The novel concept of combining many compositional spaces using graph representations to create homogeneous problem spaces, both simplifying the path planning and allowing for efficient incorporation of constraints and assumptions about problem spaces as demonstrated in Section~\ref{ssec:complexes}. 
    
\end{enumerate}

In addition to the above, three other new contributions are given in this work:

\begin{enumerate}
    
    \item Sections~\ref{sec:randomuniformsampling}~and~\ref{sec:simplexgrid} discuss random sampling and grid construction in simplex spaces in the context of the composition of chemical spaces. In the process, several theoretical results critical to the problem, which have not been discussed previously in this context, are presented. For instance, the commonly found random sampling of a $d-1$ hypercube and rejection of compositions $>100\%$ to sample a $d$-component space, commonly found in software, has a rejection rate exhibiting factorial growth and can severely impact when deploying ML models.
    
    \item In Section~\ref{ssec:internalgrid}, a new algorithm was developed to efficiently create internal (subspace-exclusive) grids in simplex spaces based on an algorithm from the literature (modified-NEXCOM \cite{Chasalow1995AlgorithmPoints}) without introducing any overhead. It is beneficial in cases of, for instance, sampling only $d$-component materials in $d$-component chemical space without considering lower-order points.

    \item In a few areas, Section~\ref{ssec:functionallygraded} leverages its general character to go beyond the usual FGM literature introduction. For instance, it contrasts elemental spaces with attainable design spaces and discusses the use of similar compositions (alloy grades) in the design process to reduce cost and greenhouse emissions without making prior assumptions.
    
\end{enumerate}

\section*{Data Availability}

No datasets were generated or analyzed during the current study. The research presents a theoretical algorithm and mathematical equations that do not rely on empirical data.

\section*{Code Availability}

The \texttt{nimplex} software described in this work has been published as free open-source software (FOSS) under the MIT license. It can be effortlessly used as a \emph{native} Nim library, \emph{native} compiled Python library, or Command Line Interface (CLI) tool interfacing with nearly any language through binary data or plain text.

All distributions of the source contain (1) the core library, (2) additional utilities, (3) testing procedures, (4) use examples, (5) quick-start guide using Python/CLI in the form of a Jupyter notebook, (6) \texttt{devcontainer.json} specification, and (7) documentation. They are available through:

\begin{itemize}
    \item The documentation page at \href{https://nimplex.phaseslab.org}{nimplex.phaseslab.org}, which contains (1) installation instructions, (2) usage instructions in Python, Nim, and CLI, and (3) Application Programming Interface (API) reference. It also links to a public GitHub repository hosting the latest code (\href{https://github.com/amkrajewski/nimplex}{github.com/amkrajewski/nimplex}) at the time of writing.


    \item (Selected Major Versions) A public repository archive on Zenodo under DOI: \href{https://doi.org/10.5281/zenodo.10611931}{10.5281/zenodo.10611931}.
\end{itemize}

\section*{Acknowledgments}

This work has been funded through grants: NSF-POSE \textbf{FAIN-2229690}, ONR \textbf{N00014-23-2721}, and DOE-ARPA-E \textbf{DE-AR0001435}. 

Adam M. Krajewski would like to thank \textbf{Gonville \& Caius College} at the University of Cambridge and \textbf{Dr. Gareth Conduit} for generously hosting him as a visiting postgraduate student during the writing of this publication, and \textbf{Peter and Carol Thrower} for sponsoring the fellowship.

We would also like to thank \textbf{Luke Myers}, \textbf{Ricardo Amaral}, and \textbf{Alexander Richter} for testing code exercises and proofreading the documentation.

\section*{Author Contributions}
\textbf{Adam M. Krajewski:} Conceptualization, Methodology, Software, Writing - Original Draft, Validation, Visualization
\textbf{Alison Beese:} Funding acquisition, Writing - Review \& Editing
\textbf{Wesley F. Reinhart:} Funding acquisition, Writing - Review \& Editing
\textbf{Zi-Kui Liu:} Funding acquisition, Supervision, Writing - Review \& Editing, Resources

\section*{Competing Interests}

The authors declare no competing interests.

\printbibliography

\appendix
\section{Appendix A} \label{app1}

As mentioned in Section~\ref{ssec:functionallygraded}, different grades of base metals may have very different costs associated with them. For instance, as of December 2023, at \href{https://www.fishersci.com}{Fisher Scientific online store (fishersci.com)}, one can purchase:
\begin{itemize}
    \item High-purity Zr wire: $250cm$ of $0.25mm$-diameter (AA00416CB) for $\$317$ or $\approx 390\frac{\$}{g}$
    \item $99.2\%$ (Zr+$4.5\%$Hf) wire: $200cm$ of $0.25mm$-diameter (AA43334G2) for $\$63$ or $\approx 100\frac{\$}{g}$. 
    \item $99.97\%$ (Hf+$3\%$Zr) wire: $200cm$ of $0.25mm$-diameter (AA10200G2) for $\$200$ or $\approx 156\frac{\$}{g}$. 
\end{itemize} 

Now, if one tries to create FGMs which navigates Zr-rich regions in Hf-containing space, there are two possible choices for Zr source, namely, pure Zr or the (Zr+$4.5\%Hf$) alloy. The first one enables all possible Zr fractions, unlike the latter which establishes the minimum Hf fraction at 4.5\% at the "Zr" corner of the attainable space tetrahedron (anonymous example of this is in Figure~\ref{fig:fgmspaces}). Such an ability may be necessary, e.g., to avoid infeasible regions of space, but if not, it represents an unnecessary cost. 

For instance, to obtain (Zr+$10\%wt$Hf) alloy, one can combine high-purity Zr and (Hf+$3\%Zr$) for $\approx 360\frac{\$}{g}$ or equivalently from (Zr+$4.5\%$Hf) and (Hf+$3\%$Zr) for $\approx 103\frac{\$}{g}$, representing 3.5 times cost reduction.

At Fisher Scientific, as of writing this, the pure-Zr wire is only available in $0.25mm$ diameter, thus, the above considerations were restricted to it to keep the comparisons fair. However, for the (Zr+$4.5\%$Hf) grade, many less-expensive form factors are available as it is much more industry relevant. Furthermore, larger package sizes ($\geq50g$) are available driving the cost down further. For instance, the following $1mm$ wires can be purchased:

\begin{itemize}
    \item $99.2\%$ (Zr+$4.5\%$Hf) wire: $10m$ of $1mm$-diameter  (AA14627H2) for $\$130$ or $\approx 2.5\frac{\$}{g}$. 
    \item $99.97\%$ (Hf+$3\%$Zr) wire: $5m$ of $1mm$-diameter (AA10205CC) for $\$580$ or $\approx 11.3\frac{\$}{g}$. 
\end{itemize}

If the above are used, one can now obtain the same (Zr+$10\%wt$Hf) alloy for $\approx 3\frac{\$}{g}$ or 120 times cheaper relative to using high purity Zr in the only available physical form factor.

\section*{Appendix B}  \label{app2}

The equation for the fraction of a cube bound by [111] plane, equivalent to the result for $f(4)$ in Section~\ref{sec:randomuniformsampling}, can be quickly obtained by considering that the volume of a pyramid is given by $V = \frac{A_B h}{3}$, where $A_B$ is the base area of equilateral triangle $\frac{\sqrt{3}}{4}\times\sqrt{2}^2 = \frac{\sqrt{3}}{2}$ and $h$ is $\frac{1}{\sqrt{3}}$. Thus we get 
$$V = \frac{\frac{\sqrt{3}}{2} \frac{1}{\sqrt{3}}}{3} = \frac{1}{6}$$ 
agreeing with the aforementioned result in Section~\ref{sec:randomuniformsampling}.

\section*{Appendix C}  \label{app3}

The equation for "forward" and "backward" jumps in 3-simplex graph corresponding to a quaternary chemical system. 

\begin{minted}{nim}
proc neighborsLink4C(...): void =
  let jump0 = 1  #binom(x, 0)=1
  let jump1 = binom(1+ndiv-x[0]-x[1], 1)
  let jump2 = binom(2+ndiv-x[0], 2)
  
  if x[0] != 0:
    # quaternary
    neighbors[i].add(i - jump2)
    # quaternary
    neighbors[i].add(i - jump2 - jump1)
    # quaternary
    neighbors[i].add(i - jump2 - jump1 - jump0) 
  
  if x[1] != 0:
    # ternary
    neighbors[i].add(i - jump1)    
    # ternary
    neighbors[i].add(i - jump1 - jump0)         
    # quaternary
    neighbors[i].add(i + jump2 - jump1 - x[1])  
  
  if x[2] != 0:
    # binary
    neighbors[i].add(i - jump0)     
    # ternary
    neighbors[i].add(i + jump1 - jump0)    
    # quaternary
    neighbors[i].add(i + jump2 - jump0 - x[1])  
  
  if x[3] != 0:
    # binary
    neighbors[i].add(i + jump0)        
    # ternary
    neighbors[i].add(i + jump1)          
    # quaternary     
    neighbors[i].add(i + jump2 - x[1])              
\end{minted}

\end{document}